\documentclass[10pt,amssymb,superscriptaddress,showkeys,twocolumn,prx]{revtex4-2}
\usepackage{bm}
\usepackage{graphicx}
\usepackage{dcolumn}
\usepackage{bm}
\usepackage{changes}
\usepackage{amsmath,amssymb}
\usepackage{titlesec}
\usepackage{lineno}
\usepackage{hyperref}




\makeatletter
\renewcommand{\@caption@fignum@sep}{\space} 
\makeatother
\titleformat{\section}[block]{\normalfont\bfseries\large}{\thesection}{1em}{}
\titlespacing*{\section}{0pt}{\baselineskip}{0.5\baselineskip}

\titleformat{\subsection}[block]{\normalfont\bfseries\small}{\thesubsection}{1em}{}
\titlespacing*{\subsection}{0pt}{\baselineskip}{0.5\baselineskip}

\hypersetup{
 colorlinks=true,
 linkcolor=blue,
 filecolor=blue,  
 urlcolor=blue,
 citecolor=blue,
}

\begin{document}
\title{Observation of nonlinear higher-order topological insulators with unconventional boundary truncations}

\author{Changming Huang}
\thanks{These authors contributed equally}
\affiliation{Department of Physics, Changzhi University, Changzhi, Shanxi 046011, China}
\author{Alexander V. Kireev}
\thanks{These authors contributed equally}
\affiliation{Institute of Spectroscopy, Russian Academy of Sciences, Troitsk, Moscow, 108840, Russia}

\author{Yuxin Jiang}
\thanks{These authors contributed equally}
\affiliation{Aerospace Information Research Institute, Chinese Academy of Sciences, Beijing 100094, China}
\affiliation{College of Precision Instrument and Optoelectronics
Engineering, Tianjin University, Tianjin 300072, China. }

\author{Victor O. Kompanets}
	\affiliation{Institute of Spectroscopy, Russian Academy of Sciences, Troitsk, Moscow, 108840, Russia}%
	
\author{Ce Shang}
\email{shangce@aircas.ac.cn}
\affiliation{Aerospace Information Research Institute, Chinese Academy of Sciences, Beijing 100094, China}
\affiliation{College of Precision Instrument and Optoelectronics
Engineering, Tianjin University, Tianjin 300072, China. }
\author{Yaroslav V. Kartashov}
\email{kartashov@isan.troitsk.ru}
\affiliation{Institute of Spectroscopy, Russian Academy of Sciences, Troitsk, Moscow, 108840, Russia}%
\author{Sergei A. Zhuravitskii}
	\affiliation{Institute of Spectroscopy, Russian Academy of Sciences, Troitsk, Moscow, 108840, Russia}%
	\affiliation{Quantum Technology Centre, Faculty of Physics, M. V. Lomonosov Moscow State University, Moscow, 119991, Russia}
\author{Nikolay N. Skryabin}
\affiliation{Institute of Spectroscopy, Russian Academy of Sciences, Troitsk, Moscow, 108840, Russia}%
\affiliation{Quantum Technology Centre, Faculty of Physics, M. V. Lomonosov Moscow State University, Moscow, 119991, Russia}%
\author{Ivan V. Dyakonov}
\affiliation{Quantum Technology Centre, Faculty of Physics, M. V. Lomonosov Moscow State University, Moscow, 119991, Russia}%
	
\author{Alexander A. Kalinkin}
\affiliation{Institute of Spectroscopy, Russian Academy of Sciences, Troitsk, Moscow, 108840, Russia}%
\affiliation{Quantum Technology Centre, Faculty of Physics, M. V. Lomonosov Moscow State University, Moscow, 119991, Russia}		
\author{Sergei P. Kulik}
\affiliation{Quantum Technology Centre, Faculty of Physics, M. V. Lomonosov Moscow State University, Moscow, 119991, Russia}%

\author{Fangwei Ye}
\email{fangweiye@sjtu.edu.cn}
\affiliation{School of Physics and Astronomy, Shanghai Jiao Tong University, Shanghai 200240, China}	
\affiliation{School of Physics, Chengdu University of Technology, Chengdu, 610059, China }

\author{Victor N. Zadkov}
\affiliation{Institute of Spectroscopy, Russian Academy of Sciences, Troitsk, Moscow, 108840, Russia}%
\affiliation{Department of Physics, Higher School of Economics, Moscow, Russia, 105066, Russia}%

\date{\today}

\begin{abstract}
In higher-order topological insulators (HOTIs), topologically nontrivial phases are usually associated with the shift of Wannier centers to topologically nontrivial positions on the edges of the unit cells, and the emergence of fractional spectral charges in the corners of the lattice upon its truncation that keeps the number of its unit cells integer.  Here we propose theoretically and illustrate \textcolor{black}{experimentally a different approach to the construction of HOTIs}. This approach utilizes lattices with incomplete unit cells and achieves localized modes of topological origin across a broader parameter space. When truncation disrupts translational symmetry by cutting through the interior of multiple unit cells, boundary modes in our system emerge for both trivial and topologically nontrivial positions of the Wannier centers. We link these modes to the appearance of \textcolor{black}{fractional Wannier centers}. We also demonstrate that linear boundary states give rise to rich families of stable solitons bifurcating from them in the presence of focusing nonlinearity. \textcolor{black}{Multiple types of thresholdless topological solitons} with different internal symmetries are observed in waveguide arrays with triangular configurations featuring incomplete unit cells for any dimerization of waveguide spacings. Our results expand the family of HOTIs and pave the way for the observation of boundary states \textcolor{black}{with different symmetries}.
\end{abstract}
                              \maketitle
\section*{\textcolor{black}{Introduction}}

\begin{figure*}[t]
\centering
\includegraphics[width=2\columnwidth]{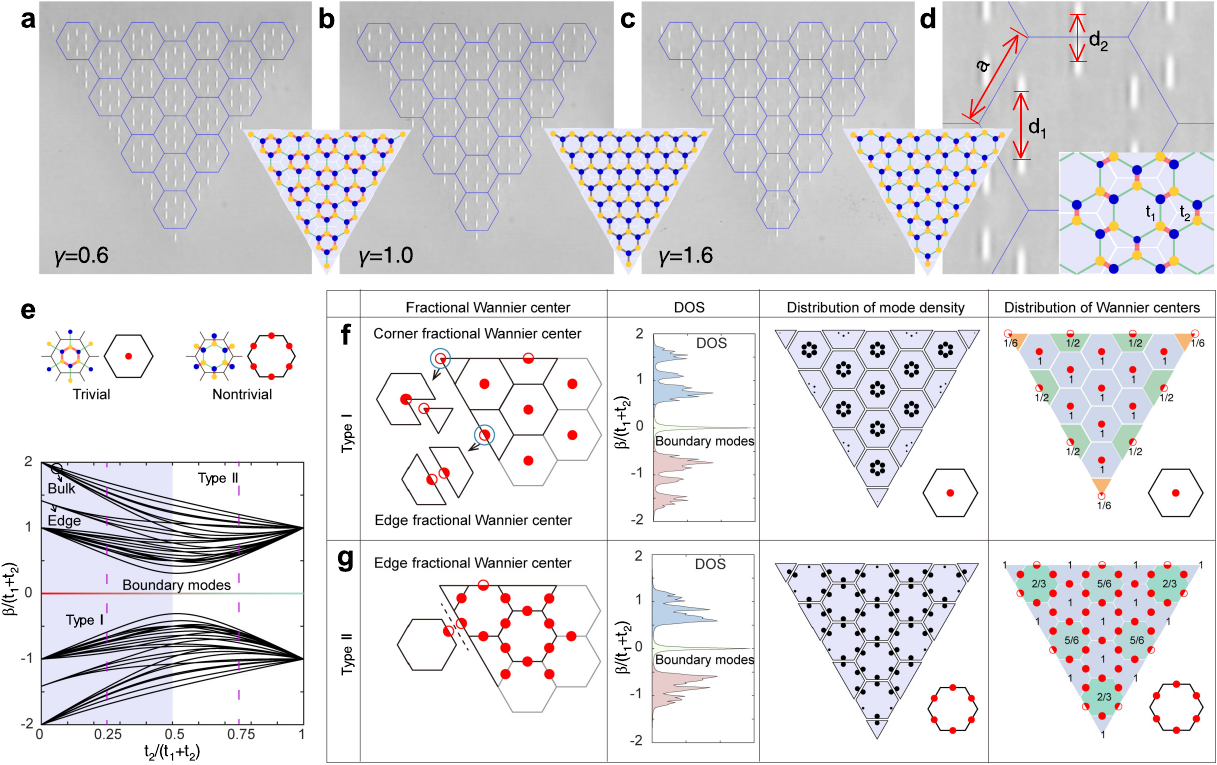}
\caption{\textbf{Construction of \textcolor{black}{higher-order topological insulators} with unconventional boundary truncations}.
Microphotographs of the triangular \textcolor{black}{femtosecond} laser-written waveguide arrays with complete and incomplete honeycomb unit cells (the former are indicated by blue contours superimposed on the photomicrographs) with different degrees of dimerization $\gamma=0.6$ (\textbf{a}), $\gamma=1.0$ (\textbf{b}) and $\gamma=1.6$ (\textbf{c}), where ${\gamma=d_1/d_2}$ defines the ratio between the distance $d_1$ between the nearest waveguides within the unit cell and the distance $d_2$ between the waveguides of neighboring cells (\textbf{d}). The insets schematically show weak and strong couplings in such arrays. (\textbf{e}) Propagation constants $\beta$ of the linear eigenmodes as a function of ${t_2}/{(t_1+t_2)}$, calculated with the tight-binding model. Note the existence of boundary states with $\beta=0$ both at $t_2<t_1$ and $t_2>t_1$. (\textbf{f}), (\textbf{g}) Schematic representation of the concept of \textcolor{black}{fractional Wannier center}, density of states, and distributions of mode density and Wannier centers within the array in type-I [$t_2<t_1$] and type-II [$t_2>t_1$] phases of the system, shown with different background colors in (\textbf{e}).}
\label{fig1}
\end{figure*}

Topological insulators represent \textcolor{black}{a class of} materials that support excitations propagating along their boundaries, whose properties and existence are guaranteed by the nontrivial topology of the bulk \cite {RevModPhys.82.3045, RevModPhys.83.1057}. The broad class of topological insulators includes materials in which topologically non-trivial phases arise due to the breaking of certain symmetries of the system, including time-reversal and inversion symmetries, or due to specific deformations of the underlying lattice structure and dynamical modulation of the system parameters \cite{wang2009observation, Yang2019, ren2023observation, Zhang2023elight, PhysRevLett.132.176302, PhysRevLett.133.236603}. To date, topological insulators have been demonstrated in solid-state physics \cite{Chang2013}, acoustics \cite{lin2023topological}, and mechanics \cite{Ssstrunk2015}, the physics of cold atoms and matter waves \cite{RevModPhys.91.015005}, in various optoelectronic and photonic systems \cite{RevModPhys.91.015006,zhang2023}. Photonic topological insulators extend the concept of electronic topological insulators to the electromagnetic domain and offer unique possibilities for controlling the structure of optical fields, their ultrafast switching, routing, and diffraction control \cite{Rechtsman2013, Noh2018, pyrialakos2022, Fritzsche2024, Huang2024, Flower2024}. Robust in-gap edge states in such materials protected by the topology of the system can resist disorder and defects, making them ideal for the design of topologically protected transmission and lasing devices. The central concept in the theory of topological insulators is that the symmetry of the lattice and the geometry of its boundaries are key to determining the properties, domains of existence, and dimensionality of topological boundary states. A conventional $d$-dimensional topological insulator supports $(d-1)$-dimensional topological states at its boundaries.

Recently, \textcolor{black}{a class of higher-order topological} insulators (HOTIs) was introduced. Such systems support nontrivial topological boundary modes characterized by a codimension of at least two \cite{Benalcazar61, PhysRevB.96.245115, Peterson2018, xie2021}. Thus, $d$-dimensional HOTI of order $n$ supports $(d-n)$-dimensional localized states at its boundaries, and these states also enjoy topological protection against disorder and defects. Symmetry plays a pivotal role in the study of these topological states, and symmetry-based analysis resulted in the prediction and classification of a multitude of crystalline insulators \cite{PhysRevLett.106.106802, Schindler2018sc}, see classification in \cite{Bradlyn2017, PhysRevB.99.245151}. Photonic HOTIs based on lattices with different symmetries have also been reported \cite{ezawa2018higher, el2019corner, Mittal2019, chen2019direct, xie2019visualization, Guo2020, liu2021, wang2021higher, kirsch2021nonlinear, arkhipova2023observation, zhong2024observation, Hu2024}. Usually, topologically nontrivial phases in crystalline HOTIs are associated with shifts of the Wannier centers to topologically nontrivial positions within the unit cells of periodic lattice, while truncation of the lattice that keeps the number of the unit cells in its integer may lead to filling anomaly for certain types of truncation and appearance of fractional spectral charges, indicating the existence of localized corner states in cells characterized by such charges. The emergence of higher-order corner states in such systems is expected only in certain configurations, tightly connected with non-trivial positions of the Wannier centers.

In this paper, we present a more general configuration of photonic HOTI with unconventional boundary truncations that result in the intersection of the interior of several unit cells, leading to the coexistence of complete and incomplete cells at the boundary. To this end, we use the \textcolor{black}{femtosecond} (fs) laser writing technique to generate waveguide arrays with a desired internal structure and external geometry. Remarkably, the nontrivial topological properties in our HOTIs emerge not from the intrinsic structure of the material, as in electronic topological insulators, but from the waveguide array inscribed in it, where the key factors are the geometric arrangement of the waveguides, the coupling strengths, and the boundary truncation. We show that a boundary state protection mechanism still exists in this system, which can be associated with \textcolor{black}{fractional Wannier centers}, and that it supports localized states of topological origin in a much larger parameter range compared to standard HOTIs (including in the regime where the latter become trivial insulators). Furthermore, we exploit the fact that our experimental system, unlike electronic and acoustic topological systems, possesses a nonlinear response that is crucial for controlling the localization and propagation dynamics of topological excitations \cite{zhang2018resonant, xue2018, maczewsky2020nonlinearity, arkhipova2022observation}, the realization of topological lasers \cite{harari2018topological, bandres2018topological}, the generation of higher harmonics \cite{You2020, Kruk2021} and topological solitons \cite{PhysRevLett.117.143901, kartashov2016modulational, mukherjee2020sol, mukherjee2021sol, Veenstra2024} and experimentally observe the formation of a variety of stable nonlinear higher-order topological states with different symmetries and stability properties bifurcating from linear corner and edge modes. Our results extend the understanding of HOTIs and \textcolor{black}{provide avenues} for the observation and characterization of topological states in such systems.

\section*{\textcolor{black}{Results and Discussion}}

The appearance of the topological phase in crystalline HOTIs is closely related to the structure of Wannier functions - a set of orthogonal functions that provides a convenient representation of the eigenstates of the crystalline system \cite{RevModPhys.84.1419}. Wannier centers, as central points of maximally localized Wannier functions, are a representation of mode densities in real space protected by crystalline symmetry \cite{PhysRevB.101.115115, Liu2021bulk}. For a Wannier center in a topologically trivial position, the associated mode density is restricted to a single unit cell (no fractional charge can arise if the lattice is truncated to obtain an integer number of unit cells). For a Wannier center in a topologically non-trivial position, the associated mode density is evenly distributed over neighboring unit cells (fractional charge occurs when the lattice is truncated). The occurrence of fractional charge can be visualized by integrating the local density of states in the occupied band per unit cell. A fractional charge is a signature of the “filling anomaly” in the terminology of solid-state physics. The theory in \cite{Peterson2020, Peterson2021, shang2024} analyzes the locations of the Wannier centers for different system parameters and explains the topological origin of HOTIs via the fractional charges. Thus, in crystalline HOTIs containing an integer number of unit cells, the transition to a topologically nontrivial phase can be controlled by introducing dimerization into the intracell ($t_1$) and intercell ($t_2$) coupling strengths, resulting in a topological phase with corner modes at $t_2>t_1$ and a trivial phase corresponding to $t_2<t_1$.

In contrast, here we present a HOTI with unconventional edge truncation that generates a structure with multiple \textcolor{black}{complete} and \textcolor{black}{incomplete} unit cells and show that the higher-order edge state protection mechanism exists in it in both $t_2>t_1$ and $t_2<t_1$ regimes. Our HOTI is based on a honeycomb waveguide array with a triangular configuration created using the fs laser writing technique \cite{Rechtsman2013, kirsch2021nonlinear}. To control the coupling strengths $t_{1,2}$ (which determine the structure of the eigenmodes of the system), we adjust the position of the waveguides within each unit cell by changing the ratio $\gamma=d_1/d_2$ (dimerization parameter) of the waveguide spacing within the cell $d_1$ and between the cells $d_2$. Microphotographs of the arrays with $\gamma<1$ (in this case $t_2<t_1$), $\gamma=1$ ($t_2=t_1$), and $\gamma>1$ (in this case $t_2>t_1$) are shown in Fig.~\ref{fig1}\textbf{a}, \textbf{b} and \textbf{c}, while Fig.~\ref{fig1}\textbf{d} illustrates the notations. The blue lines in Fig.~\ref{fig1}\textbf{a}, \textbf{b}, and \textbf{c} mark complete unit cells of the structure. As can be seen, the waveguides in the corners and some waveguides at the edge belong to incomplete cells.

\begin{figure*}[!hpt]
\centering
\includegraphics[width=1.75\columnwidth]{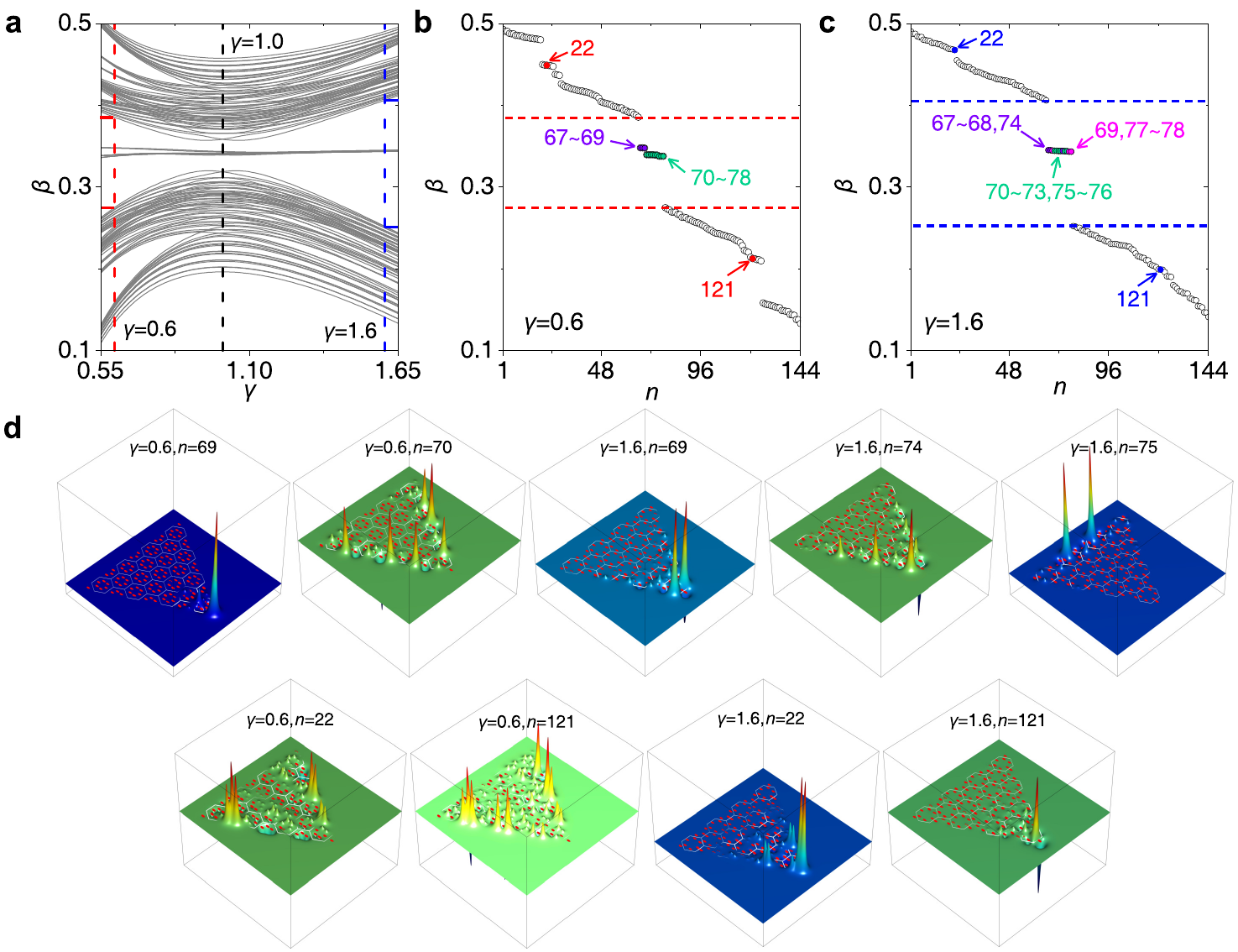}
\caption{\textbf{Linear spectrum of the system and its eigenmodes.}
Linear spectrum of the array showing the eigenvalues of all linear modes as a function of the dimerization parameter $\gamma$ \textcolor{black}{(\textbf{a})} as well as the eigenvalues at $\gamma=0.6$ \textcolor{black}{(\textbf{b})} and $\gamma=1.6$ (\textbf{c}), which correspond to the red and blue dashed lines in the $\beta(\gamma)$ dependence, respectively. The top line in (\textbf{d}) shows representative linear modes $\psi$ with eigenvalues that fall into the gap at $\gamma=0.6$ and $1.6$. At $\gamma=0.6$, the modes with indices $n=67\sim69$ correspond to the corner modes, while the modes with $n=70\sim78$ are edge modes. For $\gamma=1.6$, the indices $n=67\sim 69, 74, 77, 78$ correspond to the corner modes (including the modes $n=67$, 68, and 74 with two out-of-phase peaks, while the modes $n=69$, 77 and 78 have two in-phase peaks near the corner), while the indices $n=70\sim73$, 75, and 76 correspond to the edge modes. The bottom row in (\textbf{d}) shows linear modes from two bands with indices $n=22$ and 121, which are localized at $\gamma=1.6$ (i.e., represent \textcolor{black}{bound states in the continuum}) and delocalized at $\gamma=0.6$. The field distributions of all linear modes in the spectral gap can be found in \textcolor{black}{Supplementary Figure S1}.}\label{fig2}
\end{figure*}

Figure \ref{fig1}\textbf{e} shows the eigenvalues $\beta$ of the linear modes of such an array with $81$ sites as a function of $t_2/(t_1+t_2)$. Here the eigenvalues of array modes were calculated using the standard tight-binding Hamiltonian $H$ (\textcolor{black}{see the tight-binding approximate model in Supplementary Note 1}), which only considers the couplings between the nearest neighbors, and takes into account that all waveguides are shallow and support only single mode defined by amplitude $a_{m,n}$, where $m,n$ is the waveguide index in two-dimensional honeycomb array (in contrast to tight-binding models in electronic systems that usually apply the concepts of multiple orbitals on a single site \cite{chen2024}. Unlike conventional HOTIs on honeycomb arrays, where the topological phase only occurs at $t_2>t_1$, in our system, boundary states are found \textcolor{black}{both} at $t_2<t_1$ (i.e.~$\gamma<1$) and $t_2>t_1$ (i.e.~$\gamma>1$). There is still a transition point $t_2=t_1$, which distinguishes between type I and type II phases, which are shown in Fig. \ref{fig1}\textbf{e} with different colors (in a very large array, the gap would close at this point, while in our system it remains open due to its finite size). We also study two special cases $t_1=3t_2$ in the type-I phase and $t_2=3t_1$ in the type-II phase, as indicated by the dashed lines in Fig.~\ref{fig1}\textbf{e}. Since both phases (see Fig. \ref{fig1}\textbf{f} and Fig. \ref{fig1}\textbf{g}) support coexisting corner and edge states in the gap, as confirmed by the calculation of the density of states ${\rm{DOS}}(\beta)=-\lim_{\delta \rightarrow 0}{\rm{Im}}[{\rm{Tr}}(\beta-H+i\delta)^{-1}]$, this implies that even if the truncation occurs through the interior of several unit cells, there is still a topological mechanism supporting the existence of localized modes. This also implies that the Wannier centers, which are usually considered as a whole, can be decomposed into parts, i.e., the topological properties can be explained by the concept of \textcolor{black}{fractional Wannier center}. The left images in Fig.~\ref{fig1}\textbf{f} and \ref{fig1}\textbf{g} show the positions of the Wannier centers (red circles) within the unit cells. In the type I phase ($\gamma<1$), the Wannier centers are located at so-called topologically trivial positions (reflecting the positions of strong coupling links), in the centers of the cells, while truncation leads to the appearance of fractionated Wannier centers in the corners and at the edges of the lattice (Fig.~\ref{fig1}\textbf{f}). In the type II phase ($\gamma>1$), the Wannier centers are located at so-called topologically non-trivial positions (again reflecting the positions of strong coupling links), at the edges of the cells, and truncation of the array cutting some of these centers now leads to the appearance of fractional Wannier centers (partly white, partly red circles) at the edges (Fig.~\ref{fig1}\textbf{g}), but within other unit cells compared to the type I case. The analysis of the eigenmodes allows the general conclusion that topologically non-trivial localized boundary modes (edge or corner modes) only occur where the truncation passes through the Wannier centers, i.e., their occurrence is directly related to the occurrence of the \textcolor{black}{fractional Wannier centers}. Thus, Fig.~\ref{fig1} illustrates that the key difference between the $\gamma<1$ and $\gamma>1$ phases lies in the positions of the Wannier centers, which lead to different topological origins of boundary modes: For $\gamma<1$ both corner and edge states arise from fractional Wannier centers located within \textcolor{black}{incomplete} unit cells, resulting in unconventional topological states, while for $\gamma>1$ they correspond to conventional topological states similar to those that arise in conventional higher-order insulators and associated with Wannier centers near the boundaries of \textcolor{black}{{complete} unit cells}.

Although conventional topological invariants that rely on symmetry are poorly defined in this context, we can define the fractional spectral charge \( Q_b \) corresponding to the boundary cell $b$:
\begin{equation}
 Q_b = \sum_{m}\sum_{(x,y)\in b}\left|\left\langle\psi_m \right|P_{x,y}\left|\psi_m\right\rangle\right|\quad(\bmod 1),
\end{equation}
where \( P_{x,y} = |x,y\rangle\langle x,y| \) represents the position operator and \( \left| \psi_m \right\rangle \) denotes the \(m\)-th eigenmode for the corresponding occupied band \cite{PhysRevLett.133.233804}. To generalize the concept of fractional charge, we consider the incomplete unit cells as if they were complete, with completely filled additional imaginary vacant areas. The spectral charges defined according to this procedure are shown in the right panels in Fig.~\ref{fig1}\textbf{f} and \ref{fig1}\textbf{g}. In Fig.~\ref{fig1}\textbf{f} the fractional charges for orange (corner) zones and green (edge) zones are given by $1/6$ and $1/2$, respectively. In Fig.~\ref{fig1}\textbf{g}, the fractional charges for green edge zones are equal to $5/6$, while for green cells near the corners, they are equal to $2/3$ (to define corresponding spectral charges we take into account that Wannier centers shared between cells contribute charge $1/2$ to each cell, except for Wannier centers that fall onto outer boundaries, and normalize the charge for all cells such that bulk cells possess spectral charge $1$). They immediately allow the identification of incomplete (in the type I phase) or complete (in the type II phase) cells in which the charge is fractional and localized boundary modes can arise. It can also be seen that such cells always contain \textcolor{black}{fractional Wannier centers}.



\begin{figure*}[t]
\centering
\includegraphics[width=1.75\columnwidth]{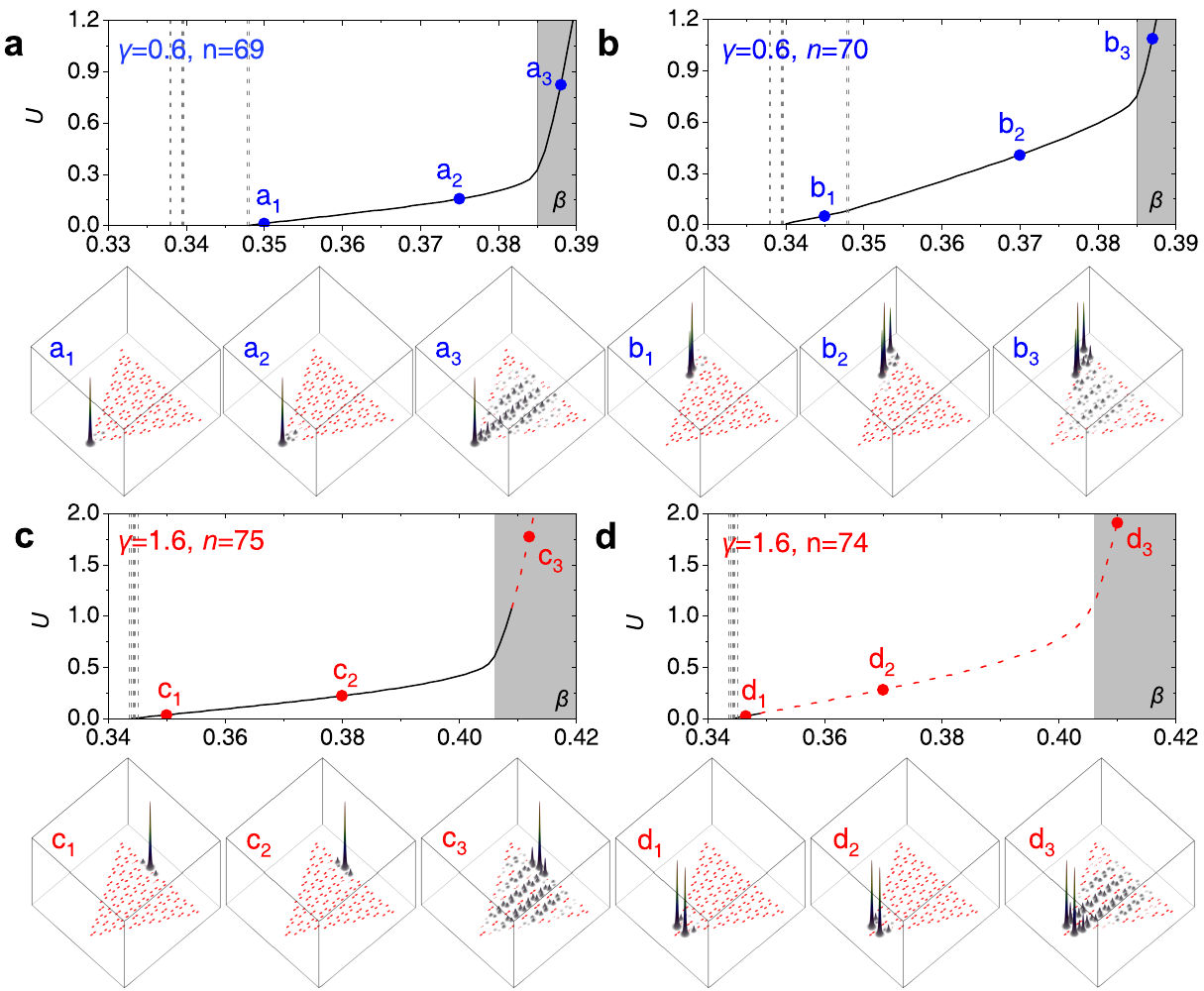}
\caption{\textbf{Families of thresholdless topological solitons arising at different \textcolor{black}{dimerization values $\gamma$.}} 
\textcolor{black}{Soliton power $U$ versus propagation constant $b$, illustrating} soliton families bifurcating from different linear localized modes: (\textbf{a}) the corner mode with $n=69$ at $\gamma=0.6$; (\textbf{b}) the edge mode with $n=70$, which is located in the incomplete edge cell at $\gamma=0.6$; (\textbf{c}) the edge mode with $n=75$, which is located in the complete cell at $\gamma=1.6$; and (\textbf{d}) the out-of-phase mode near the corner with $n=74$ at $\gamma=1.6$. The bottom line shows the intensity distributions $|\psi|^2$ corresponding to the points in the $U(\beta)$ dependencies. Stable branches are shown with solid black lines, while unstable branches are shown with red dashed lines. Gray regions indicate bulk bands, and vertical gray dashed lines represent the eigenvalues of the linear modes.}\label{fig3}
\end{figure*}

\begin{figure*}[t]
\centering
\includegraphics[width=2\columnwidth]{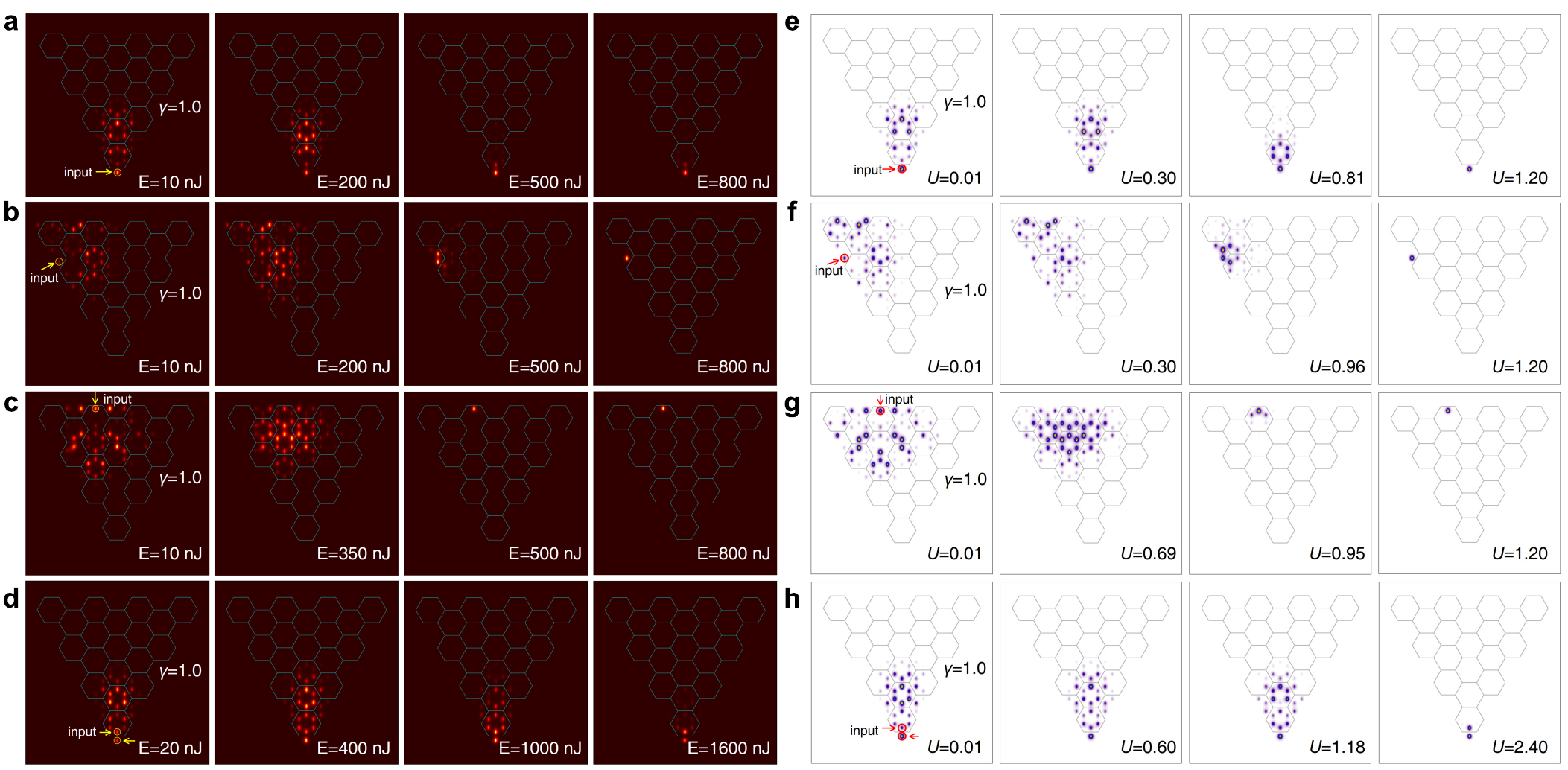}
\caption{\textbf{Excitation of linear and non-linear states at \textcolor{black}{dimerization parameter} $\gamma=1.0$.}
Comparison of the experimental output intensity distributions [(\textbf{a})-(\textbf{d}), maroon background] with theoretically calculated output distributions [(\textbf{e})-(\textbf{h}), white background] for different excitation positions indicated by the arrow and the circles. The contours superimposed on the intensity distributions show only complete (not truncated) unit cells of the lattice. \textcolor{black}{(\textbf{a,e})} Waveguide in the bottom corner; \textcolor{black}{(\textbf{b,f})} the seventh waveguide from the top left corner at the left edge; \textcolor{black}{(\textbf{c,g})} the ninth waveguide from the top left corner at the top edge; and \textcolor{black}{(\textbf{d,h})} two waveguides in the bottom corner, out-of-phase excitation. The total input pulse energies in the experiment and the input powers in the simulations are given on each panel. Here and below, the corresponding $E$ values are typically doubled in the case of two-site excitation to obtain the same nonlinear contribution to the refractive index in each waveguide.}\label{fig4}
\end{figure*}

\begin{figure*}[t]
\centering
\includegraphics[width=2\columnwidth]{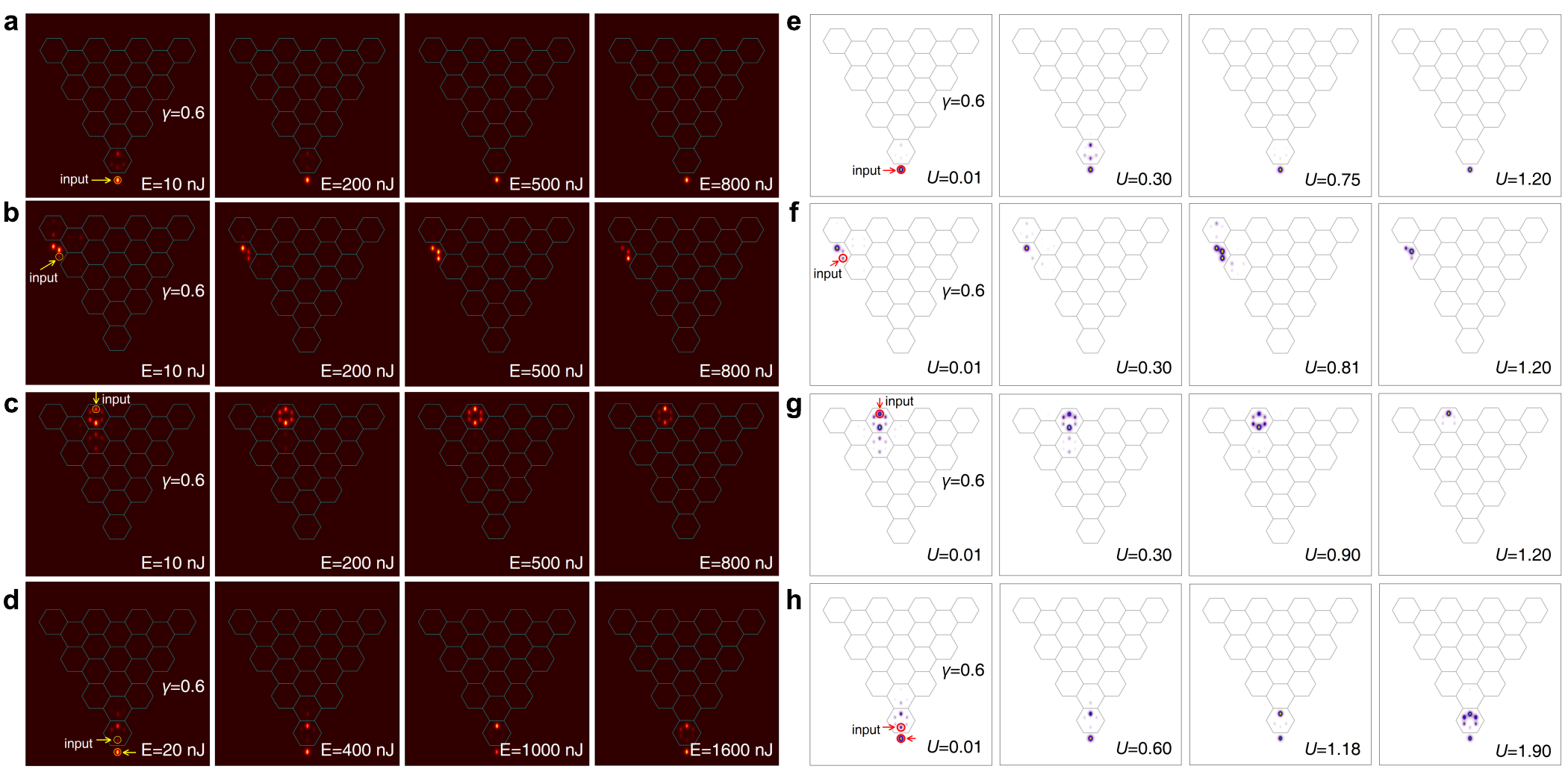}
\caption{\textbf{Excitation of linear and nonlinear states at \textcolor{black}{dimerization parameter} $\gamma=0.6$.}
The arrangement and meaning of the panels are as in Fig. 4. \textcolor{black}{(\textbf{a}) and (\textbf{e}) show} the intensity distributions for the single-site excitation of the waveguide in the lower corner (where the mode with index $n=69$ is located); \textcolor{black}{(\textbf{b}) and (\textbf{f}) show} the single-site excitation of the seventh waveguide from the upper left corner at the left edge (strong overlap with the edge mode with $n=72$ in the incomplete cell); \textcolor{black}{(\textbf{c,g})} single-site excitation of the ninth waveguide at the upper edge belonging to the complete unit cell; \textcolor{black}{(\textbf{d,h})} out-of-phase excitation of two lower waveguides.}\label{fig5}
\end{figure*}

\begin{figure*}[t]
\centering
\includegraphics[width=2\columnwidth]{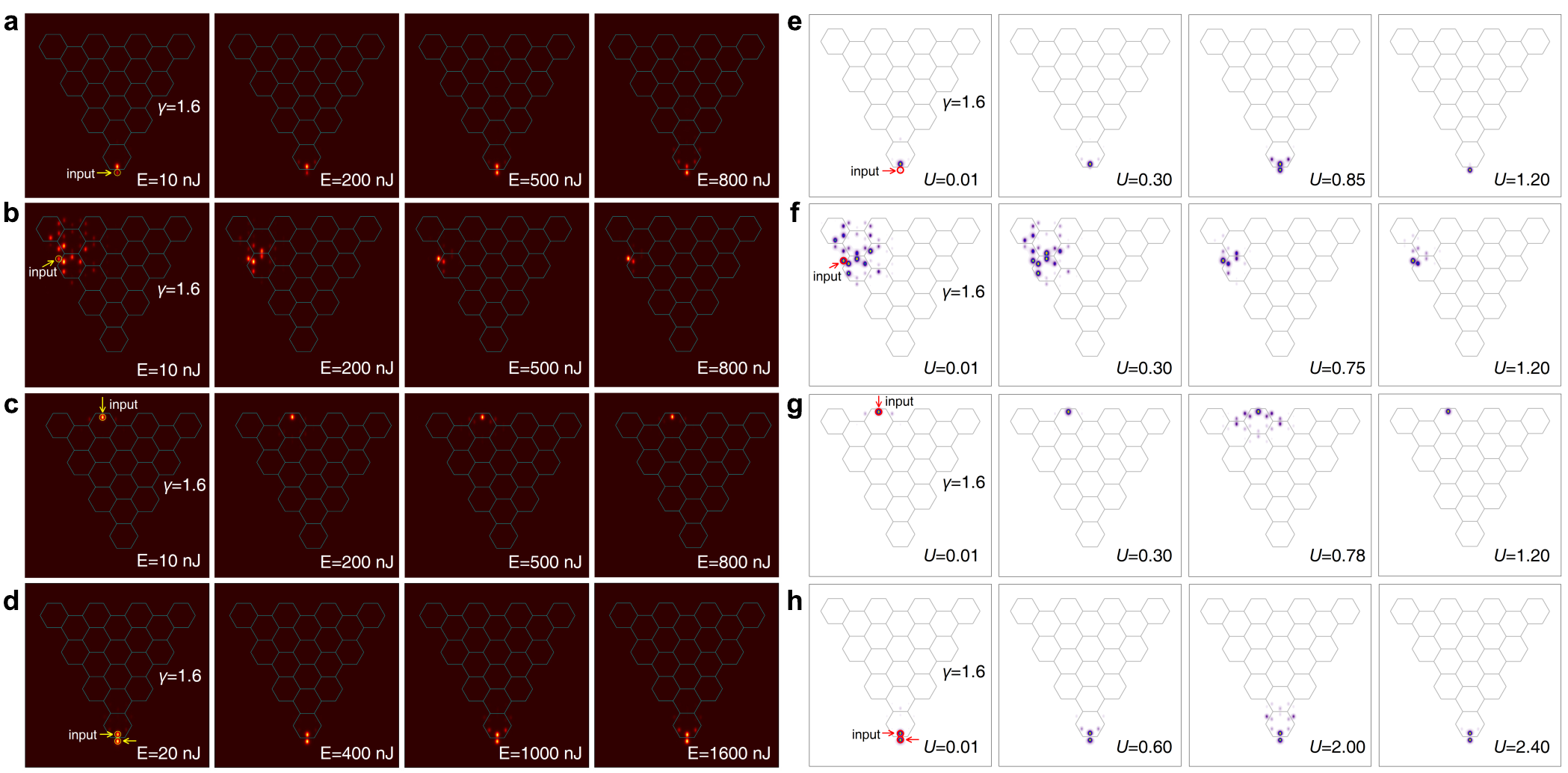}
\caption{\textbf{Excitation of linear and non-linear states at \textcolor{black}{dimerization parameter} $\gamma=1.6$.}
The arrangement and meaning of the panels are as in Fig. 5. \textcolor{black}{(\textbf{a,e})} Single-site excitation of the waveguide in the lower corner waveguide; \textcolor{black}{(\textbf{b,f})} single-site excitation of the seventh waveguide from the upper left corner at the left edge; \textcolor{black}{(\textbf{c,g})} single-site excitation of the ninth waveguide at the upper edge (overlapping with the edge mode with \textcolor{black}{index} $n=75$ in the whole unit cell); \textcolor{black}{(\textbf{d,h})} phase-shifted excitation of two lower waveguides overlapping with the phase-shifted corner mode with $n=121$.}\label{fig6}
\end{figure*}

\subsection*{Linear modes and soliton families}

To confirm the above conclusions based on the tight-binding model and to demonstrate the appearance of new types of topological boundary modes in HOTI arising from unconventional truncation, we further use the continuous dimensionless nonlinear Schr\"odinger equation describing the paraxial propagation of light along the $z$-axis:
\begin{equation}\label{eq2}
i\frac{\partial \Psi}{\partial z}=-\frac{1}{2}\nabla^2\Psi-\mathcal{V}(x,y)\Psi-|\Psi|^2\Psi,
\end{equation}
where $\Psi(x,y,z)$ is the field amplitude, $z$ is the propagation distance, $x$ and $y$ are the transverse coordinates and $\nabla=(\partial/\partial x,\partial/\partial y)$ is the transverse Laplacian (for the normalization of all quantities see \textbf{Methods}). The function $\mathcal{V}(x,y)$ describes the refractive index distribution inside the array with a honeycomb inner structure and triangular outer geometry consisting of Gaussian waveguides:
\begin{align*}
\mathcal{V}(x, y)=p \sum_{m, n} e^{-\left(x-x_{m,n}\right)^{2} / w_{x}^{2}-\left(y-y_{m,n}\right)^{2} / w_{y}^{2}}. 
\end{align*}
Here, the parameter $p$ represents the array depth proportional to the refractive index contrast $\delta n$ within the structure, the positions of the waveguide centers ($x_{m,n},y_{m,n}$) define a honeycomb array with introduced Kekulé distortion, whose strength is characterized by the ratio $\gamma=d_1/d_2$ of the intra- and intercell distances of the waveguides [see Fig.~\ref{fig1}\textbf{d}], where $\gamma=1$ corresponds to an undistorted structure. The side length of the unit cell is denoted by $a$. In addition, the waveguides are elliptical with widths $w_y>w_x$ due to the writing process. In accordance with the actual parameters of the laser-written structures, the dimensionless parameters are set as $a=5.72$, $w_x=0.25$, $w_y=0.75$, and $p=4.7$. The proposed HOTIs are composed of $144$ waveguides forming $19$ complete unit cells and several incomplete cells. Microphotographs in Fig.~\ref{fig1}\textbf{a}-\textbf{c} show three typical waveguide arrays with different $\gamma$ values written by the fs laser. Note that the continuous model takes into account all details of the refractive index landscape and thus accounts for the coupling between all waveguides in the system, which is an advantage over the tight-binding description.

To characterize the topological properties of this structure, it is crucial to determine its linear spectrum, which we achieve by searching for linear eigenmodes $\Psi(x,y,z)=\psi(x,y)e^{i\beta z}$, where $\psi(x,y)$ is the modal field and $\beta$ is the eigenvalue (propagation constant). The eigenmodes result from the subsequent linear eigenvalue problem:
\begin{equation}\label{eq3}
\beta \psi=\frac{1}{2}\nabla^2\psi+\mathcal{V}(x,y)\psi,
\end{equation}
solved using the finite-difference method.

The linear spectrum of our array is shown in Figs.~\ref{fig2}\textbf{a}-\textbf{c}, while typical modal profiles are shown in Fig.~\ref{fig2}\textbf{d}. The most representative feature of the spectrum is the presence of localized boundary modes of topological origin in the spectral gap for \textcolor{black}{all values} of the dimerization parameter $\gamma$ (except for the point $\gamma=1$), see Fig.~\ref{fig2}\textcolor{black}{\textbf{a}}. This is in sharp contrast to conventional HOTIs, where localized states only occur for $\gamma>1$. Note that for very large structures with incomplete cells at the edge, the gap closes at $\gamma=1$, whereas in Fig.~\ref{fig2}\textcolor{black}{\textbf{a}} it only remains open due to the finite size effect, since we model the structure of the same size as in the experiment (where certain technological constraints exist that limit the size of the lattice that can be inscribed). From Fig.~\ref{fig2}\textcolor{black}{\textbf{b}} and \ref{fig2}\textbf{c} it can be seen that there are $12$ corner or edge modes within the spectral gap for $\gamma<1$ and for $\gamma>1$. Among these modes, corner modes appear in triplets that become degenerate for sufficiently small or large $\gamma$ values (see, for example, modes $n=67\sim69$ at $\gamma=0.6$). Since these modes are degenerate, one can use their linear combinations to construct states residing in only one, two, or three corners of the array. It is noteworthy that the internal structure of the corner modes at $\gamma<1$ and $\gamma>1$ is qualitatively different, as their primary maxima are located in different waveguides (compare the corner modes in Fig.~\ref{fig2}\textbf{d}, top row). Such modes gradually delocalize when $\gamma\to 1$. In addition, the system is characterized by the existence of bound states in the continuum (BICs) in the spectral band, which only occur for sufficiently large values of $\gamma$ (see e.g. the blue arrows in Fig.~\ref{fig2}\textbf{c}, which show the location of such states at $\gamma=1.6$ in the spectrum, and the typical profiles in Fig.~\ref{fig2}\textbf{d}, bottom row). A detailed description of the modes within the spectral gap of the array with unconventional boundary can be found in \textcolor{black}{Supplementary Figure S1 and Supplementary Note 2}. It should be also stressed that emergence of strongly localized corner modes, like mode $n=69$ at $\gamma=0.6$, or BICs $n=22$ and $n=121$ at $\gamma=1.6$, cannot be explained as a result of hybridization of conventional edge states because all these corner modes have considerable intensity in a waveguide from incomplete cell in the corner and disappear if this waveguide is removed.

For the sake of comparison, in \textcolor{black}{Supplementary Figure S2} we show the linear spectrum and representative eigenmodes of the array with the same parameters, but with usual truncation along the boundaries of the unit cells, which keeps the integer number of the unit cells. By comparing this linear spectrum with the spectrum of the array with unconventional truncation presented in Fig.~\ref{fig2}\textcolor{black}{\textbf{a}}, one arrives at the following conclusions. In striking contrast to unconventionally truncated HOTIs (i) no localized modes emerge when $\gamma<1.0$, (ii) at $\gamma>1.0$ strongly localized on two outermost waveguides modes (akin to modes $n=22$ and $n=121$ in HOTI with unconventional truncation) disappear from the spectrum, and (iii) the spread of eigenvalues of the modes residing in the spectral gap of array with usual truncation at $\gamma>1.0$ increases considerably, while corner and edge modes in the latter structure acquire more complex and broad shapes with multiple peaks having comparable amplitudes. Thus, one can conclude that unconventional truncation substantially changes the spectrum of the system and structure of modes (\textcolor{black}{also see relevant discussion in Supplementary Note 3}).

It should be stressed that the arrays considered in this work are sufficiently large and finite-size effects that could lead to coupling between localized states in different corners are not visible in them (in practice such coupling occurs only when modes strongly expand as $\gamma \to 1$), so that significant reduction of the size of the structure would be needed for observation of such effects. Moreover, being states of topological origin, corner modes persist in our case even in the presence of disorder, as long as the spectral gap, where they emerge, remains open. To illustrate this, we show the transformation of the linear spectrum of the array with unconventional boundary truncation for increasing disorder levels in 
\textcolor{black}{Supplementary Figure S3}. The gap shows the tendency for reduction with the increase of disorder level (as observed for both diagonal and off-diagonal disorder), in the case of off-diagonal, is more pronounced at $\gamma<1$ due to the way in which we set the disorder. Nevertheless, corner modes persist in the gap and are not destroyed until disorder becomes sufficiently strong to close it completely. For $\gamma>1$ we also observe remarkable insensitivity of the eigenvalues of corner and edge modes to disorder (\textcolor{black}{see Supplementary Note 4}).

Since our system possesses focusing nonlinearity, we try to observe solitons of topological origin bifurcating from the above states in the spectral gap. We consider solitons of the form $\Psi(x,y,z)=\psi(x,y)e^{i\beta z}$, where $\beta$ is now an independent parameter that determines the soliton profile $\psi$ and its power $U=\iint|\psi(x,y)|^2dxdy$. If we insert this expression into Eq.~(\ref{eq2}), we obtain the equation $(1/2)\nabla^2\psi+\mathcal{V}(x,y)\psi-\beta \psi+|\psi|^2\psi=0$, which was solved using the Newton iteration method. In Fig.~\ref{fig3}, we show representative examples of soliton families $U(\beta)$ and profiles of solitons of four different types. Because all such solitons bifurcate from linear edge or corner modes, their power $U$ vanishes when the soliton propagation constant $\beta$ approaches the eigenvalue of the linear state, from which bifurcation occurs. This means that all such solitons can form even at low powers, i.e., they are \textcolor{black}{thresholdless}, since they do not require a minimal power threshold for their formation, as it happens for usual two-dimensional lattice solitons. At $\gamma<1$, we present two different families of solitons bifurcating from the corner mode and the edge mode in the incomplete unit cell, see Fig.~\ref{fig3}\textbf{a} and \ref{fig3}\textbf{b}. As the power $U$ increases, the propagation constant $\beta$ of the solitons shifts upwards from the linear eigenvalue towards the bulk band. Most of the power of the solitons bifurcating from the corner mode is concentrated in the corner waveguide, while a small part of the power is localized in the adjacent cell (see panel $\textbf{a}_1$ in Fig.~\ref{fig3}\textbf{a}). When $U$ increases, the soliton gradually widens (see panel $\textbf{a}_2$), and when $\beta$ enters the band, it spreads over the entire array (see panel $\textbf{a}_3$). The behavior of soliton bifurcating from the edge mode is somewhat different - because the propagation constant of edge mode is lower than that of corner state, the family of edge solitons emanating from edge mode remains well-localized at the edge (see panel $\textbf{b}_1$ in Fig.~\ref{fig3}\textbf{b}) only until $\beta$ crosses the eigenvalue of corner state. The crossing leads to a coupling of the edge soliton with the corner state and a progressive increase of the intensity in the corner (see the panels $\textbf{b}_2$ and $\textbf{b}_3$ in Fig.~\ref{fig3}\textbf{b}), followed by a coupling with the bulk modes when $\beta$ enters the band. Remarkably, both corner and edge solitons are dynamically stable in their entire existence region, even when they couple with bulk modes (this means the input perturbations imposed on such solitons do not grow upon propagation over huge distances $z$ exceeding sample length by orders of magnitude). The stability of the soliton families was checked using linear stability analysis, the details of which are summarized in \textbf{Methods}.

For $\gamma>1$, a representative family of solitons bifurcating from the edge mode in the complete unit cell at the boundary is shown in Fig.~\ref{fig3}\textbf{c}. Such solitons are typically strongly confined to three waveguides of the complete unit cell for the propagation constants $\beta$ in the gap (see panels $\textbf{c}_1$ and $\textbf{c}_2$ in Fig.~\ref{fig3}\textbf{c}) and extend into the bulk only at sufficiently high powers (see panel $\textbf{c}_3$ in Fig.~\ref{fig3}\textbf{c}). As the linear stability analysis shows, this nonlinear family is completely stable in the gap [see solid part of the $U(\beta)$ curve indicating the stable branch], but destabilizes within the band [see dashed part of the $U(\beta)$ curve]. A representative example of a corner soliton family at $\gamma>1$ bifurcating from out-of-phase linear corner mode is shown in Fig.~\ref{fig3}\textbf{d}. The transformation of shapes of such solitons with the increase of $\beta$ is illustrated in panels $\textbf{d}_1$-$\textbf{d}_3$ of Fig.~\ref{fig3}\textbf{d}. Interestingly, this solution becomes unstable shortly after the bifurcation from the linear corner mode (see the dashed part of the family in Fig.~\ref{fig3}\textbf{d}). This instability is analogous to classical symmetry-breaking instability that in a finite gap occurs for a corner soliton with out-of-phase spots. It is a result of bifurcation from the present anti-symmetric family of stable asymmetric corner soliton family characterized by larger intensity in one of the near-corner waveguides (not shown in Fig.~\ref{fig3}\textbf{d}). These results show that topological solitons in HOTIs with unconventional truncations occur in a variety of forms with different internal structures and for any value of the dimerization parameter $\gamma$. Several different stable soliton branches can coexist in the same gap.

It should be stressed that the corner solitons studied here are robust nonlinear states that persist in the presence of disorder in the underlying array. As long as the spectral gap remains open and a linear corner or edge mode persists in the spectral gap, this mode gives rise to the family of solitons bifurcating from it. Such soliton families obtained in arrays with different realizations and levels of the off-diagonal disorder are presented in 
\textcolor{black}{Supplementary Figure S4} for both $\gamma=0.6$ and $\gamma=1.6$ values of dimerization parameter. Remarkably, the studied corner and edge soliton families remained stable within the spectral gap despite the presence of disorder in the array (\textcolor{black}{see relevant discussion in Supplementary Note 4}).

\subsection*{Experimental observation of the nonlinear states}

To verify the existence of such linear localized modes and solitons bifurcating from them for each value of the dimerization parameter $\gamma$ in an array with unconventional boundary truncation, we used the fs laser direct writing technique to inscribe such arrays with triangular shapes in $10$ cm long fused silica slabs. A detailed description of the writing parameters can be found in the \textbf{Methods} section. For excitation, we used $280~\textrm{fs}$ high-intensity laser pulses at a wavelength of $\lambda=800~\textrm{nm}$ with an energy $E$ per pulse originating from a $1~\textrm{kHz}$ Ti:sapphire laser system that can be focused into selected waveguides. The input peak power of such pulses can be evaluated as a ratio of the pulse energy to its $280~\textrm{fs}$ duration, so that ${1\,\rm nJ}$ of input pulse energy can be equated to ${2.5\,\rm kW}$ of peak power. The error in measured pulse energy and evaluated peak power does not exceed $5 \%$ and is mainly connected with fluctuations of energy of pulses, accuracy of determination of their duration, and focusing efficiency into waveguides. The dispersion for such pulses at $10~\textrm{cm}$ can be neglected, allowing us to consider only the spatial dynamics in Eq. (\ref{eq2}). Nevertheless, one should take into account that measured output intensity distributions represent averaged patterns containing contributions from regions around the pulse peak propagating in a strongly nonlinear regime, and contributions from linearly diffracting pulse tails. For this reason, experimental patterns in the figures below may appear somewhat more extended than corresponding theoretical output intensity distributions obtained from a purely spatial model (\ref{eq2}). For two-site excitations with a controllable phase difference between two beams, the Michelson interferometer scheme was employed (see \textbf{Methods}).

First, we examined the excitation dynamics in an array with a regular honeycomb structure, without introducing Kekulé distortion into it, i.e.,~for $d_1=d_2$ and $\gamma=1.0$. Fig.~\ref{fig4} compares experimentally recorded output intensity distributions (\textbf{a}-\textbf{d}) for four different and typical types of excitations of the edge and corner waveguides of the array with results of numerical modeling in the frames of Eq.~(\ref{eq2}) (\textbf{e}-\textbf{h}). Due to the absence of localized modes in the array at $\gamma=1$, the beams in linear (pulse energy $E =10~\textrm{nJ}$) and weakly nonlinear regime (pulse energy $E =200~\textrm{nJ}$) exhibit dramatic broadening across the array irrespectively of the position of the input excitation. This is observed in the first two columns of Figs.~\ref{fig4}\textbf{a-c} and~\ref{fig4}\textbf{e-h}. With an increase of pulse energy to the values about $E =500~\textrm{nJ}$ the diffraction is considerably suppressed in some cases, and signatures of nonlinearity-induced localization start to appear in other cases (third column). For all types of excitation pulse energy of $E =800~\textrm{nJ}$ was sufficient for confinement of light practically in one waveguide at the edge or in the corner (fourth column in Figs.~\ref{fig4}a-c and~\ref{fig4}\textbf{e-g}). This is a clear signature of the formation of a nontopological soliton existing in a semi-infinite gap and requiring a considerable power threshold for its excitation. The threshold was largest for the case of out-of-phase excitation of two waveguides in the bottom corner. These experimental results are in excellent agreement with the results of theoretical simulations, illustrating that the system is indeed in a trivial phase at $\gamma=1$.

The arrays with $\gamma<1$ and $\gamma>1$ support localized topological boundary modes, but their structure is significantly different, as shown above, so that qualitatively different propagation scenarios can be expected in these cases. In Fig.~\ref{fig5} we show experimental results for the same excitation positions, but in an array with $\gamma=0.6$. In this configuration, even at low energy inputs $E =10~\textrm{nJ}$, we observe the excitation of the topological corner mode in the bottom corner (panels \textbf{a} and \textbf{e}), which corresponds to the linear mode with index $n=69$, and the edge mode (panels \textbf{b} and \textbf{f}), which is associated with the state with index $n=73$ (\textcolor{black}{see distributions in Supplementary Figure S1\textbf{a}}). Note that in the latter case, the single input beam excites a combination of several localized edge modes. For this reason, one observes light beating between three waveguides in the excited incomplete cell, but no diffraction into the bulk. Note that both these types of localized modes form on waveguides from incomplete cells. In contrast, when a corner or edge waveguide is excited within a complete unit cell, strong diffraction into the bulk is observed (see panels \textbf{c} and \textbf{g}), since such cells do not support localized topological modes for $\gamma<1$. When two bottom waveguides are excited (panels \textbf{d} and \textbf{h}), one observes a mixed scenario, where light remains in the outermost waveguide, but leaves the excited waveguide just above it (this dynamic persists even in the nonlinear regime). As the pulse energy increases, localization of the soliton in the bottom corner increases as well (panels \textbf{a} and \textbf{e}). Edge mode in a nonlinear regime demonstrates beatings depending on pulse energy $E$ with output remaining within the incomplete cell at the edge (panels \textbf{b} and \textbf{f}). For excitation of the complete unit cell on the edge, even pulse energy $E =800~\textrm{nJ}$ was not sufficient to obtain good localization (panels \textbf{c} and \textbf{g}).

Figure~\ref{fig6} compares the experimental and theoretical excitation dynamics for the dimerization parameter $\gamma=1.6$. As the investigation of the linear spectrum shows (see Fig.~\ref{fig2}), for this value of $\gamma$, there are no topological modes in the spectral gap that would be predominantly confined in a single outer corner waveguide (see \textcolor{black}{Supplementary Figure S1}). For this reason, the excitation of the corresponding corner (belonging to the incomplete cell) leads to a dynamical change of the pattern with the increase of the pulse energy $E$ (see Fig.~\ref{fig6}\textbf{a} and \ref{fig6}\textbf{e}), quite in contrast to the strong localization observed for such an input at $\gamma=0.6$. Remarkably, in this case, light switches mainly between two waveguides, the corner one and the waveguide adjacent to it, but there is no diffraction into the bulk. This is a clear indication of the excitation of the linear combination of two BICs associated with linear in-band modes with indices $n=121$ and $n=22$ (see Fig.~\ref{fig2}\textbf{d}, bottom row), localized at $\gamma=1.6$ and characterized by two out-of-phase and in-phase maxima in the two outermost corner waveguides. While single-site excitation leads to nearly equal efficiency of excitation of these two modes, which results in beatings, using two out-of-phase input beams has enabled very clean excitation of the nonlinear corner state of this type, illustrated in Figs.~\ref{fig6}\textbf{d} and \ref{fig6}\textbf{h}. It should be emphasized that despite the presumed location of this state in the band, increasing the input pulse energy did not lead to radiation into the bulk, as can be seen by comparing the output patterns at $E =400~\textrm{nJ}$ and $E =1600~\textrm{nJ}$. In contrast to the $\gamma=0.6$ case, for $\gamma=1.6$ bright spots in edge modes are located in waveguides belonging to the complete unit cells. Thus, when a waveguide in the incomplete cell is excited, one observes strong diffraction (see Figs.~\ref{fig6}\textbf{b} and \ref{fig6}\textbf{f}) that is arrested only at the highest pulse energy $E =800~\textrm{nJ}$ used here for single-waveguide inputs. In clear contrast, similar excitation of a waveguide in the complete unit cell on the edge (see Figs.~\ref{fig6}\textbf{c} and~\ref{fig6}\textbf{g}) results in the formation of a thresholdless topological edge soliton that remains well-confined at all pulse energies. 

Here, we have discussed and presented experimental evidence for the formation of only some types of topological edge and corner solitons from a variety of nonlinear states possible in this topological system. Additional experimental results illustrating the excitation of other types of solitons (\textcolor{black}{see relevant discussions in Supplementary Note 5 and Supplementary Note 6}), such as two-spot states bifurcating from the topological corner mode $n=69$ at $\gamma=1.6$, are presented in \textcolor{black}{Supplementary Figure S6}. These results show far-reaching perspectives for the realization of states of topological origin with different symmetries, which are opened up by unconventional edge truncations in HOTIs.

\section*{Conclusions}

We have proposed an approach for the realization of HOTI with unconventional edge truncations, which allows the observation of linear modes and solitons of topological origin with new types of symmetry arising for values of dimerization parameters that are usually considered topologically trivial (when Wannier centers are located in the center of the unit cell). We have proposed an explanation for the topological origin of these edge modes based on the concept of \textcolor{black}{fractional Wannier centers}. Thus, unconventional boundary truncation not only allows us to obtain localized corner modes at any value of the dimerization parameter except for $\gamma=1$, but it also leads to the appearance of corner states with new structure and generally better localization, associated with fractional spectral charges (i.e., fractional mode densities) on corresponding cells, in comparison with modes in HOTIs with usual truncation. We have also shown that in this relatively simple waveguiding system, linear boundary states lead to coexisting rich families of stable topological solitons bifurcating from linear states. Being thresholdless, such solitons inherit the internal structure of linear topological corner modes, from which they bifurcate, and therefore their variety is richer in arrays with unconventional boundary truncations in comparison with usual HOTIs. Our results expand the category of topological insulators and pave the way for their realization in other areas of science beyond optics, including cold atoms \cite{jotzu2014experimental,liang2024}, Bose-Einstein condensates \cite{RevModPhys.82.3045, RevModPhys.83.1057}, acoustics \cite{PhysRevLett.114.114301, PhysRevLett.132.236401}, and polariton condensates \cite{Wu2023, Bennenhei2024, Peng2024}.

\section*{Methods}

\subsection*{Normalization of parameters in theoretical model}

The dimensionless Eq.~(\ref{eq2}) in the main text is derived from the following dimensional version:
\begin{equation}\label{eq4}
i \frac{\partial \mathcal{E}}{\partial Z} = -\frac{1}{2k} \left( \frac{\partial^2}{\partial X^2} + \frac{\partial^2}{\partial Y^2} \right) \mathcal{E} - \frac{k}{n_0}(\delta n + n_2|\mathcal{E}|^2) \mathcal{E},
\end{equation}
where $X,Y$ are dimensional transverse coordinates and $Z$ is the propagation distance. We use the standard ``soliton'' units and introduce dimensionless transverse coordinates as ${x=X/r_0}$, ${y=Y/r_0}$, where ${r_0=10~\mu \textrm{m}}$ is the characteristic transverse scale, which also determines the relationship between the real propagation distance $Z$ and the dimensionless propagation distance as ${Z=z_d z}$, where the diffraction length ${z_d=kr_0^2\approx1.14\,\textrm{mm}}$. The real field amplitude is related to the dimensionless amplitude as ${\mathcal{E}=(n_0/k^2 r_0^2 n_2)^{1/2} \psi}$. Where ${k=2\pi n_0/\lambda}$ is the wavenumber in the medium with undisturbed refractive index $n_0$ (for fused silica $n_0\approx 1.45$ and the non-linear refractive index ${n_2\approx 2.7\times 10^{-20}\, \rm m^2/W}$), ${\lambda=800~\textrm{nm}}$ is the experimental working wavelength and $\delta n$ is the refractive index contrast that defines the structure of the shallow optical potential $\mathcal{V}(x,y)$. The dimensionless depth of this potential is given by ${p=k^2r_0^2\delta n/n_0}=4.7$, which corresponds to the refractive index contrast $\delta n \approx 5.3\times 10^{-4}$. The side length of the unit cell of the lattice in all the structures considered is ${a=5.72}$, which corresponds to ${57.2~\mu\textrm{m}}$, the waveguide widths ${w_x=0.25}$, ${w_y=0.75}$ correspond to 2.5~$\mu\textrm{m}$, 7.5~$\mu\textrm{m}$ wide elliptical waveguides, while the sample length of ${10~\textrm{cm}}$ corresponds to ${z\approx 84.3}$ (taking into account a slight reduction in length due to edge polishing). 

\subsection*{Fs-laser inscription of the waveguide arrays}
Truncated honeycomb waveguide arrays with a triangular outer geometry were fabricated using the fs laser direct writing technique. We focused laser pulses from a fiber laser system (Avesta Antaus, $515\,\textrm{nm}$ wavelength, ${1\,\textrm{MHz}}$ repetition rate, ${230\,\textrm{fs}}$ pulse duration, ${320\,\textrm{nJ}}$ pulse energy) through an aspherical lens (${\textrm{NA} = 0.3}$) into the bulk of fused silica glass (JGS1, ${10\,\textrm{cm}}$ long). To create a waveguide, the glass sample was translated relative to the waist of the writing beam with a scanning velocity of ${1\,\textrm{mm/s}}$ using a high-precision positioner (Aerotech FiberGlide 3D). This procedure was repeated to create the array in the desired configuration with complete and incomplete unit cells.

\subsection*{Experimental excitation of the waveguide arrays}

In our experimental setup, we excited one or two waveguides with femtosecond pulses of variable energy $E$. We used the ${1\,\rm kHz}$ Ti:Sapphire CPA laser system Spitfire HP (Spectra-Physics), which delivers ultrashort pulses with a duration of ${40\,\rm fs}$, a central wavelength of ${800\,\rm nm}$ and ${40\,\rm nm}$ spectral width. Before focusing into selected waveguides of the sample, the femtosecond radiation passed through an active beam position stabilization system (Avesta), an attenuator, and was further narrowed in a $4f$ single grating pulse spectrum slicer with a variable slit. Narrowing the broad pulse spectrum to ${5\,\rm nm}$ increased its duration to ${280\,\rm fs}$ and made it possible to avoid strong spectral broadening associated with self-phase modulation in the nonlinear regime. The sample with the waveguide arrays was coupled to the focusing system using a precise 6-axis positioning system (Luminos). The output radiation distribution of the arrays was recorded with a 12MP CMOS camera Kiralux (Thorlabs). For in-phase and out-of-phase excitations of two waveguides, we used a Michelson interferometer where the phase difference between the two beams was controlled by a precise rotation of the compensation plate in one of its arms. We defined the input peak power of the pulse as the ratio of the pulse energy to its duration (which in the experiment was ${280\,\rm fs}$), so that ${1\,\rm nJ}$ of input pulse energy can be equated to ${2.5\,\rm kW}$ of peak power, taking into account the losses during beam focusing into a waveguide. For example, the maximum excitation energy of ${E = 800\,\rm nJ}$ in the experimental samples presented here corresponds to the peak power of ${2.0\,\rm MW}$.

\subsection*{Linear stability analysis of soliton families}
The analysis of the stability of soliton families in this topological system was performed not only using direct propagation of perturbed inputs but also using rigorous linear stability analysis. For this purpose, the shapes of the perturbed soliton solutions were represented in the form $\Psi(x,y,z)=[\psi(x,y)+u(x,y)e^{\delta z}+iv(x,y)e^{\delta z}]e^{i\beta z}$, where $|u,v|\ll |\psi|$ are small perturbations, while $\delta$ is the growth rate of the perturbation (which can be complex). Substituting this expression into Eq.~(\ref{eq2}) and linearizing the equation around the stationary solution $\psi(x,y)$, the following linear eigenvalue problem can be derived:
\begin{equation} \label{eq5}
    \begin{split}
        \delta u & =  -\frac{1}{2}\nabla^2v+\beta v- \psi^{2} v-\mathcal{V} v, \\
        \delta v & =  +\frac{1}{2}\nabla^2u-\beta u+3 \psi^{2} u+\mathcal{V} u.
    \end{split}
\end{equation}
By solving Eq.~(\ref{eq5}) with a linear standard eigenvalue solver, we obtain the dependencies $\delta(\beta)$ of the growth rate of the perturbation on the propagation constant $\beta$. If ${\rm Re(\delta)\le0}$ holds for all perturbation modes, then the soliton solution with a given $\beta$ is stable, while it is unstable if ${\rm Re(\delta)>0}$. Stable soliton branches are shown in Fig. \ref{fig3} with solid lines, while unstable branches are shown with dashed lines.

\section*{Acknowledgments}
This work was supported in part by research project FFUU-2024-0003 of the Institute of Spectroscopy of the Russian Academy of Sciences and by the Russian Science Foundation (grant 24-12-00167). This work was also supported by the Applied Basic Research Program of Shanxi Province (202303021211191) and Project Nos. E4BA270100, E4Z127010F, E4Z6270100, E53327020D of the Chinese Academy of Sciences.

\section*{Data Availability}
The data that support the findings of this study are available from the corresponding author upon reasonable request.

\section*{Code Availability}
The program code used in this study is available from the corresponding authors upon reasonable request.

\section*{Author contributions}
\textcolor{black}{C.H., and Y.J. performed the theoretical simulations. A.V.K., and V.O.K. designed and conducted the experiments. S.A.Z., N.N.S., I.V.D., A.A.K. fabricated the samples, C.S., Y.V.K, S.P.K., V.N.Z., and F.Y. supervised the project. All the authors contributed to the discussions of the results and the preparation of the manuscript.}

\section*{Conflict of interest}
The authors declare no competing interests.

\end{document}


\title{Supplementary Information for \\ Observation of nonlinear higher-order topological insulators with unconventional boundary truncations}

\author{Changming Huang}
\thanks{These authors contributed equally}
\affiliation{Department of Physics, Changzhi University, Changzhi, Shanxi 046011, China}
\author{Alexander V. Kireev}
\thanks{These authors contributed equally}
\affiliation{Institute of Spectroscopy, Russian Academy of Sciences, Troitsk, Moscow, 108840, Russia}

\author{Yuxin Jiang}
\thanks{These authors contributed equally}
\affiliation{Aerospace Information Research Institute, Chinese Academy of Sciences, Beijing 100094, China}
\affiliation{College of Precision Instrument and Optoelectronics
Engineering, Tianjin University, Tianjin 300072, China. }

\author{Victor O. Kompanets}
	\affiliation{Institute of Spectroscopy, Russian Academy of Sciences, Troitsk, Moscow, 108840, Russia}%
	
\author{Ce Shang}
\email{shangce@aircas.ac.cn}
\affiliation{Aerospace Information Research Institute, Chinese Academy of Sciences, Beijing 100094, China}
\affiliation{College of Precision Instrument and Optoelectronics
Engineering, Tianjin University, Tianjin 300072, China. }
\author{Yaroslav V. Kartashov}
\email{kartashov@isan.troitsk.ru}
\affiliation{Institute of Spectroscopy, Russian Academy of Sciences, Troitsk, Moscow, 108840, Russia}%
\author{Sergei A. Zhuravitskii}
	\affiliation{Institute of Spectroscopy, Russian Academy of Sciences, Troitsk, Moscow, 108840, Russia}%
	\affiliation{Quantum Technology Centre, Faculty of Physics, M. V. Lomonosov Moscow State University, Moscow, 119991, Russia}
\author{Nikolay N. Skryabin}
\affiliation{Institute of Spectroscopy, Russian Academy of Sciences, Troitsk, Moscow, 108840, Russia}%
\affiliation{Quantum Technology Centre, Faculty of Physics, M. V. Lomonosov Moscow State University, Moscow, 119991, Russia}%
\author{Ivan V. Dyakonov}
\affiliation{Quantum Technology Centre, Faculty of Physics, M. V. Lomonosov Moscow State University, Moscow, 119991, Russia}%
	
\author{Alexander A. Kalinkin}
\affiliation{Institute of Spectroscopy, Russian Academy of Sciences, Troitsk, Moscow, 108840, Russia}%
\affiliation{Quantum Technology Centre, Faculty of Physics, M. V. Lomonosov Moscow State University, Moscow, 119991, Russia}		
\author{Sergei P. Kulik}
\affiliation{Quantum Technology Centre, Faculty of Physics, M. V. Lomonosov Moscow State University, Moscow, 119991, Russia}%

\author{Fangwei Ye}
\email{fangweiye@sjtu.edu.cn}
\affiliation{School of Physics and Astronomy, Shanghai Jiao Tong University, Shanghai 200240, China}	
\affiliation{School of Physics, Chengdu University of Technology, Chengdu, 610059, China }

\author{Victor N. Zadkov}
\affiliation{Institute of Spectroscopy, Russian Academy of Sciences, Troitsk, Moscow, 108840, Russia}%
\affiliation{Department of Physics, Higher School of Economics, Moscow, Russia, 105066, Russia}%

\date{\today}

\maketitle
\setlength{\parindent}{0pt}    
\setlength{\parskip}{1em}      

In the Supplementary Information, we provide a more detailed analysis of the structure of all linear modes that form within the spectral gap of the array with unconventional boundary truncations. Additional experimental results are also included, in particular for two-site corner excitations.

\textbf{Supplementary Note 1. Tight-binding approximation of photonic lattice.}
\textcolor{black}{Assuming that light is tightly confined in single-mode waveguides, we can employ the tight-binding approximation for the description of light propagation dynamics in our structure}
\begin{equation}
  i \partial_z \Psi_i(z)=-\sum_j t_{i j}(\lambda) \Psi_j(z),
  \label{e1}
\end{equation}
\textcolor{black}{where $\Psi_n$ is the light field amplitude in the $n$-th waveguide, and $t_{i j}(\lambda)$ is the coupling constant between waveguides $i$ and $j$ at the wavelength $\lambda$. To obtain the linear spectrum of the structure, one can write the field of eigenmodes in Eq.~(\ref{e1}) as $\Psi_n(z)=\Psi_n e^{i \beta z}$. One then obtains the eigenvalue problem} 
\begin{equation}
\beta \Psi_i=\sum_j t_{i j}(\lambda) \Psi_j,
 \label{e2}
\end{equation}
\textcolor{black}{where $\beta$ plays the role of energy in the analogous Schrödinger equation $H \psi_i=\beta \psi_i$. The intracell and intercell coupling terms are defined as $t_{1}=C \mathrm{e}^{-\kappa d_1}, t_{2}=C \mathrm{e}^{-\kappa d_2}$, where $C=C(\lambda)$ and $\kappa=\kappa(\lambda)$ are the wavelength-dependent experimental parameters. The model (\ref{e2}) was used for the calculation of the spectrum presented in Fig. 1(e) in the main text.}

\begin{figure}[htbp]
\centering
\includegraphics[width=1\columnwidth]{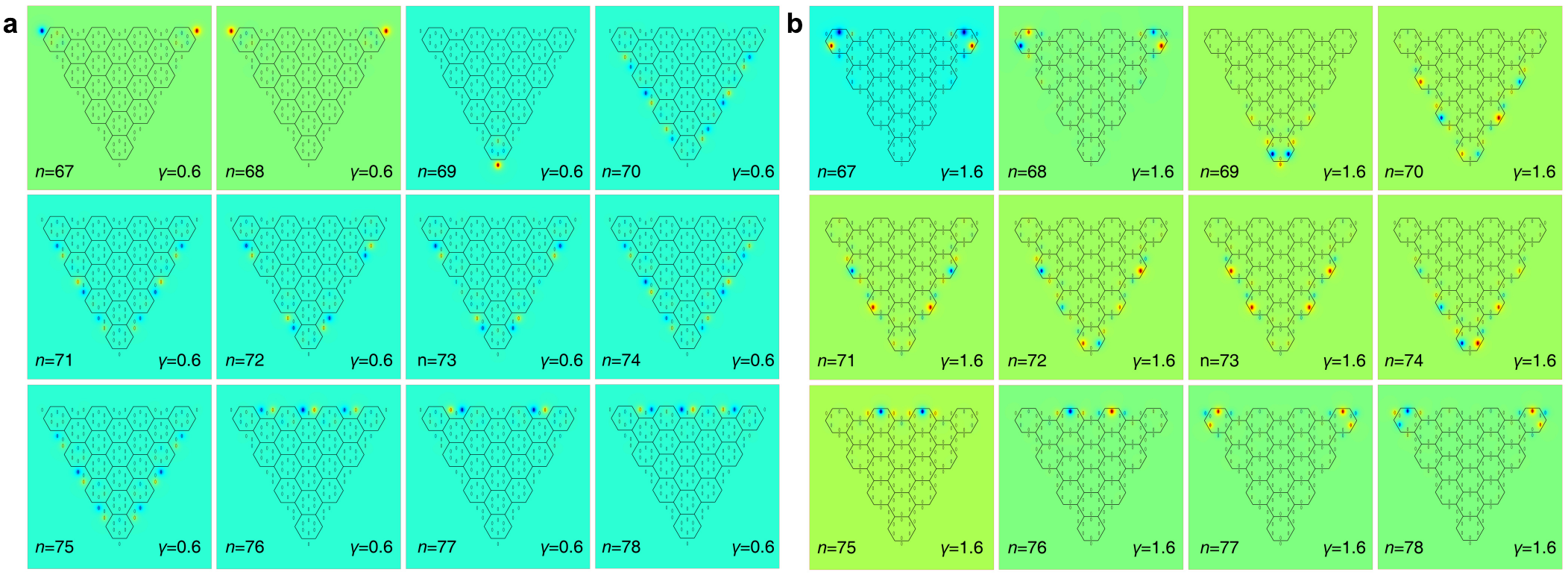}
\caption{\textbf{Field distributions in all linear eigenmodes in the spectral gap.} Field distributions in linear modes with different indices $n$ located within the spectral gap at $\gamma=0.6$ (\textbf{a}) and $\gamma=1.6$ (\textbf{b}). The shift parameter $\gamma$ and the mode indices $n$ are indicated on each panel.}
\label{figSI1}
\end{figure}

\begin{figure*}[htbp]
	\centering
	\includegraphics[width=0.85\columnwidth]{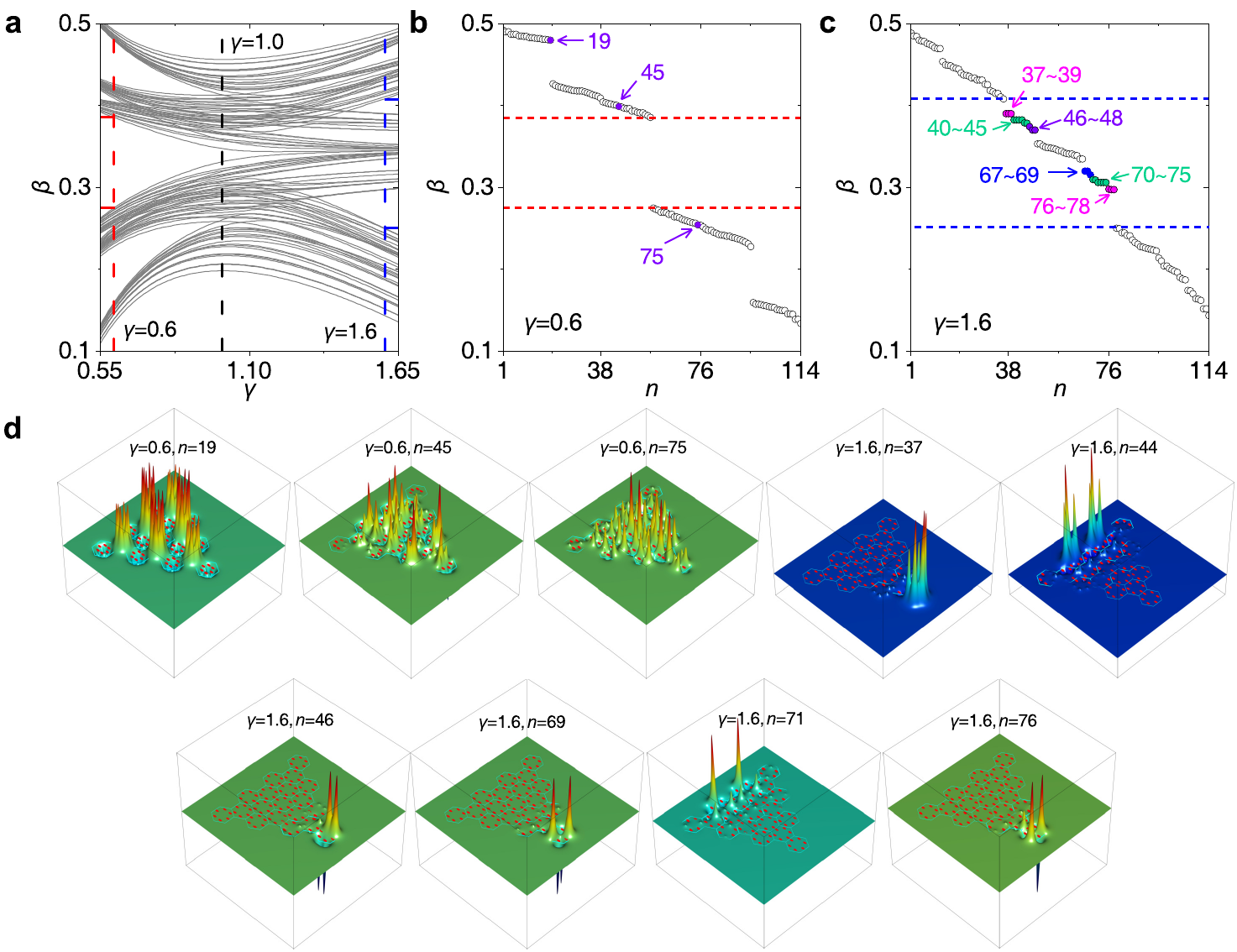}
	\caption{\textbf{Linear spectrum and eigenmodes in higher-order topological insulators with usual edge truncations}. The arrangement of panels is analogous to that in Fig.~2 in the main text. The linear spectrum of the array as a function of the dimerization parameter $\gamma$ is plotted in (\textbf{a}), while the eigenvalue distributions for $\gamma=0.6$ and $\gamma=1.6$ are displayed in (\textbf{b}) and (\textbf{c}), respectively. The three left panels in the upper row of (\textbf{d}) display representative profiles of eigenmodes at $\gamma=0.6$, while the subsequent six panels of (\textbf{d}) show representative eigenmodes for $\gamma=1.6$. At $\gamma=0.6$, all modes are delocalized. For $\gamma=1.6$, the modes with indices $n=37\sim39$, $n=46\sim48$, $n=67\sim69$ and $n=76\sim78$ correspond to the corner modes. The modes $n=40\sim45$ and $n=70\sim75$ are identified as edge modes. The former consists of representative elements with three in-phase spots in edge cells, while the latter contains elements with three out-of-phase spots in edge cells.}
	\label{figSI2}
\end{figure*}

\textbf{Supplementary Note 2. Linear modes in the gap at $\gamma=0.6$ and $\gamma=1.6$.}
To understand the properties of the linear localized modes appearing in the spectral gap of the array with unconventional boundary truncation, it is instructive to compare their field distributions. The profiles of all $12$ localized modes appearing in the spectral gap for $\gamma=0.6$ and $\gamma=1.6$ are shown in Supplementary Figure \ref{figSI1}\textbf{a} and \ref{figSI1}\textbf{b}, respectively. At $\gamma=0.6$, it is obvious that the modes $n=67\sim 69$ are the linear combinations of three degenerate corner modes centered on incomplete cells (the fact of their existence follows from the equivalence of the three corners of the structure). A linear combination of these modes can be used to generate the state localized in one or two corners. The modes with indices $n=70\sim78$ are edge modes (some of them are linear combinations of edge modes), and their common feature is that the light in them is concentrated mainly on the waveguides belonging to the incomplete unit cells (see Supplementary Figure \ref{figSI1}\textbf{a}). At $\gamma=1.6$, the modes with indices $n=67\sim 69, 74, 77, 78$ are corner modes with different internal structures (among them, the modes $n=67$, $68$ and $74$ feature two out-of-phase dominating spots in the complete unit cell near the corner, while the modes $n=69$, $77$ and $78$ feature two in-phase dominating spots in the complete cells near the corner). The modes with indices $n=70\sim73$, $75$, and $76$ are the edge modes and they exhibit bright spots mainly concentrated within complete unit cells (see Supplementary Figure \ref{figSI1}\textbf{b}). Therefore, the transition between the regimes of $\gamma<1$ and $\gamma>1$ is accompanied by a change in the locations of the intensity maxima in the localized states, and this is also accompanied by a significant change in their internal symmetry. 

\begin{figure*}[b]
	\centering
	\includegraphics[width=0.8\columnwidth]{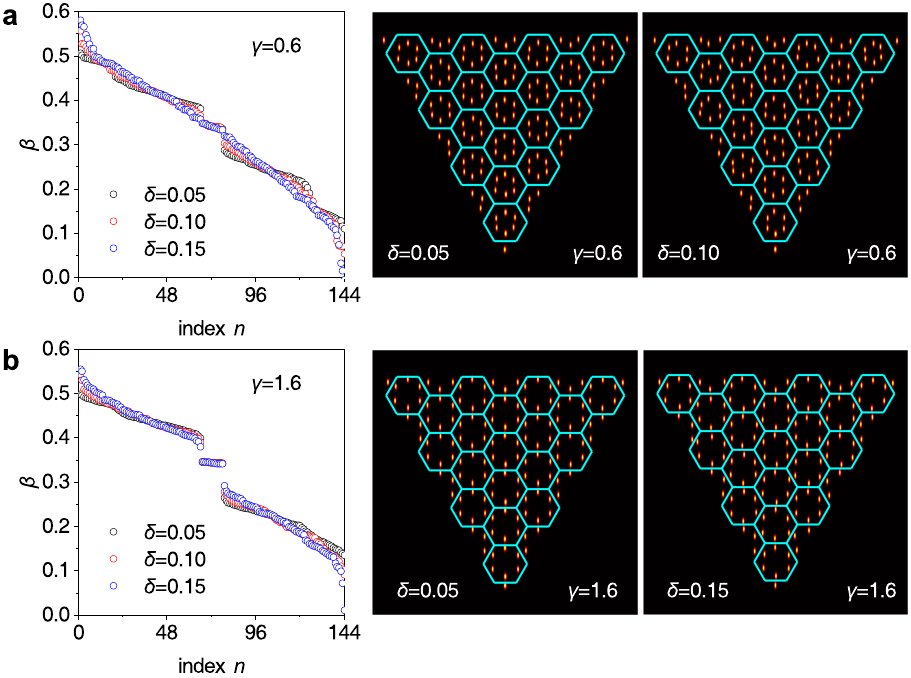}
\caption{\textbf{Linear spectra and array profiles for different disorder levels}. Panels (\textbf{a}) and (\textbf{b}) show the transformation of the linear spectrum of the array with the increase of disorder level for $\gamma=0.6$ and $\gamma=1.6$, respectively. The right-hand side of panels (\textbf{a}) and (\textbf{b}) show profiles of arrays with unconventional boundary truncation for various disorder realizations and disorder levels $\delta$.}
	\label{figSI3}
\end{figure*}

\textbf{Supplementary Note 3. Linear spectrum and eigenmodes of higher-order topological insulators (HOTIs) with usual truncation.}
\textcolor{black}{To better understand the properties and origin of linear modes in HOTI with unconventional boundary truncations, it is instructive to compare the linear spectrum of such a topological insulator with the linear spectrum of HOTI with usual truncation that occurs strictly along the edges of the unit cells. The linear spectra of HOTI with usual edge truncation are presented in Supplementary Figure \ref{figSI2}\textbf{a}-\textbf{c}, while representative eigenmodes of such a structure are shown in Supplementary Figure \ref{figSI2}\textbf{d}. In striking contrast to unconventionally truncated HOTIs, one finds that: (i) no localized modes emerge when $\gamma<1.0$, (ii) at $\gamma>1.0$ strongly localized on two outermost waveguides modes (akin to modes $n=22$ and $n=121$ in HOTI with unconventional truncation) disappear from the spectrum, (iii) the spread of eigenvalues of modes residing in the gap of HOTI with usual truncation at $\gamma>1.0$ increases considerably, while corner and edge modes in the latter structure acquire more complex and broad shapes with multiple peaks having comparable amplitudes. Thus, one can conclude that unconventional truncation substantially changes the spectrum of the system and structure of modes.}

\textbf{Supplementary Note 4. Robustness of corner states and solitons to disorder.}
\textcolor{black}{To prove the robustness of the corner modes in the presence of disorder in the underlying structure, we first analyze how disorder affects the linear spectrum of an array with unconventional boundary truncation. We first introduced position (or off-diagonal) disorder into locations of waveguide centers that are now written as $x'_{m,n}=x_{m,n}+\delta\cdot d_2 \cdot \rho_x$ and  $y'_{m,n}=y_{m,n}+\delta\cdot d_2 \cdot \rho_y$, where $x_{m,n},y_{m,n}$ are waveguide positions in regular array, $\rho_x$ and $\rho_y$ are the random numbers uniformly distributed within the interval $[-1, 1]$, $d_2=3^{1/2}a/(1+2\gamma)$ is the unperturbed inter-cell waveguide spacing, and parameter $\delta$ sets the relative strength of disorder. The examples of disordered arrays for different realizations and strengths of disorder are presented on the right side of Supplementary Figure \ref{figSI3} for two representative values of the dimerization parameter $\gamma=0.6$ and $\gamma=1.6$. Increasing disorder leads to a gradual decrease in the width of the spectral gap, where localized edge modes emerge. This effect is particularly pronounced in the $\gamma<1$ regime (see Supplementary Figure \ref{figSI3}\textbf{a}), where a smaller $\gamma$ value corresponds to stronger effective perturbation at fixed $\delta$. One can observe that for $\delta=0.15$, the gap closes; however, for lower levels of disorder, it remains open, and localized modes persist within the gap. For $\gamma>1$ the gap also shows the tendency for reduction with increase of $\delta$, but the effect is less pronounced and eigenvalues of localized modes exhibit minimal variations with disorder strength (see Supplementary Figure \ref{figSI3}\textbf{b}). A similar picture was obtained for diagonal disorder associated with fluctuations in depths $p$ of individual waveguides.}

\textcolor{black}{Once the spectral gap remains open for a given strength of disorder, the corner and edge states persist in the gap and can give rise to families of solitons. We have obtained corresponding soliton families and investigated their stability for the relative disorder strengths of $\delta=0.05$ and $0.10$, considering both (i) corner solitons at $\gamma= 0.6$ and (ii) edge solitons at $\gamma= 1.6$. Corresponding $U(\beta)$ curves and examples of soliton profiles with different powers are shown in Supplementary Figure \ref{figSI4}. As in the unperturbed array, the propagation constant of the soliton approaches the linear eigenvalue of the corresponding edge or corner mode as $U\rightarrow{0}$. This eigenvalue may be slightly shifted due to disorder in the array, but the very family of solitons persists. Linear stability analysis confirms that these solitons remain completely stable within the spectral gap, even under $10\%$ relative disorder in the waveguide array.}

\begin{figure*}[htbp]
	\centering
	\includegraphics[width=0.85\columnwidth]{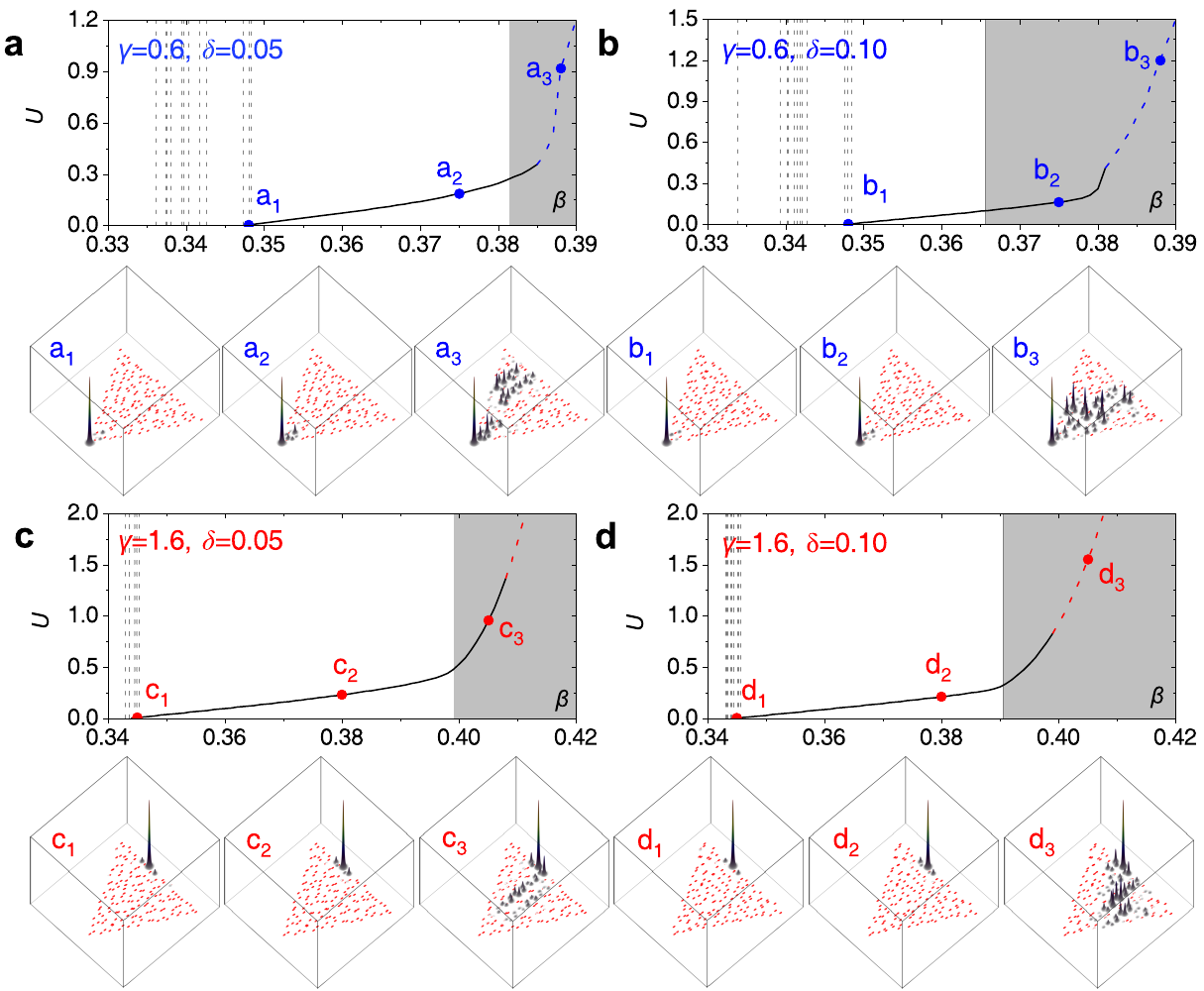}
	\caption{\textbf{Families of thresholdless topological solitons in arrays with disorder}. Families of corner solitons $U(\beta)$ bifurcating from linear corner modes at $\gamma=0.6$ exhibit $U(\beta)$ curves for disorder strengths (\textbf{a}) $\delta=0.05$ and (\textbf{b}) $\delta=0.10$. Families of edge solitons $U(\beta)$ bifurcating from linear edge modes at $\gamma=1.6$ for (\textbf{c}) $\delta=0.05$ and (\textbf{d}) $\delta=0.10$. Below each panel, we show intensity distributions $|\psi|^2$ in solitons corresponding to the points in the $U(\beta)$ dependencies. Stable branches are shown with solid black lines, while unstable branches are shown with red dashed lines. Gray regions indicate bulk bands, and vertical gray dashed lines represent the eigenvalues of the linear modes in the spectral gap.}
	\label{figSI4}
\end{figure*}

\begin{figure*}[htbp]
	\centering
	\includegraphics[width=1\columnwidth]{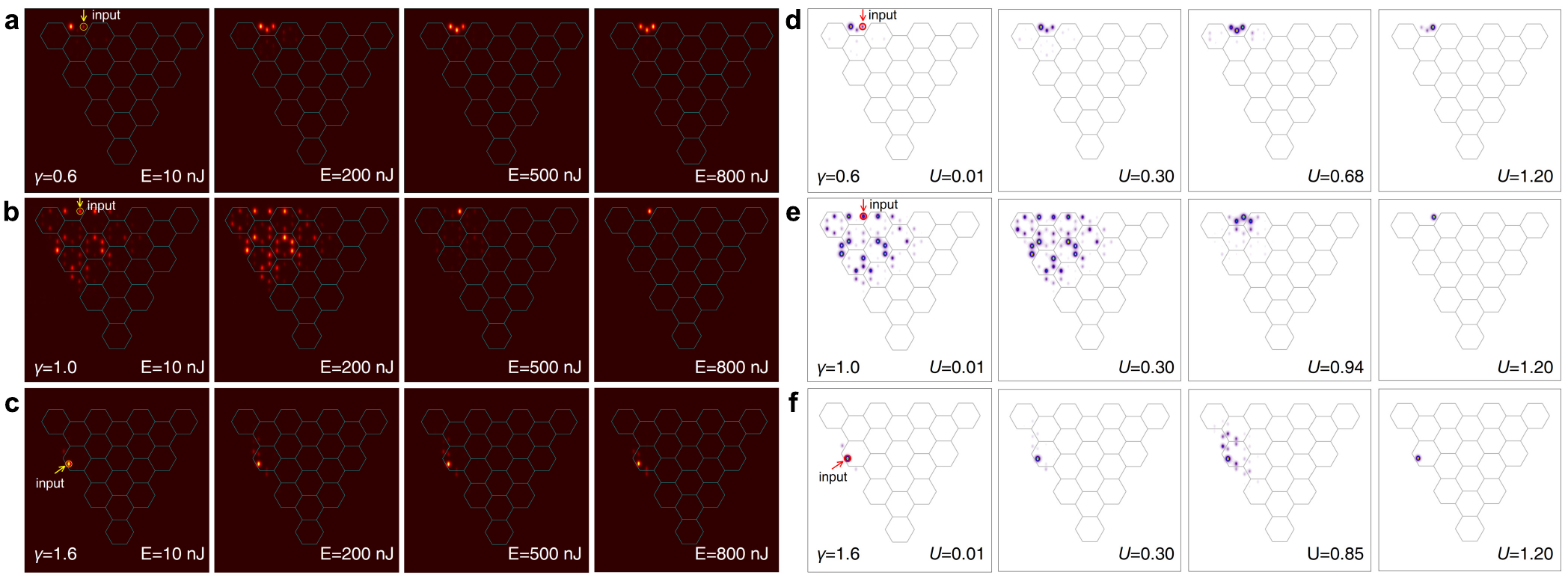}
	\caption{\textbf{Excitation of linear and nonlinear states with single-site input}. Comparison of the experimental output intensity distributions [(\textbf{a})-(\textbf{c}), maroon background] with theoretically calculated output distributions [(\textbf{d})-(\textbf{f}), white background] for different excitation positions indicated by the arrow and the circles. The contours superimposed on the intensity distributions show only complete ( non-truncated) unit cells of the lattice. (\textbf{a}) Single-site excitation of the seventh waveguide from left to right at the top edge at $\gamma=0.6$ ( overlapping with the edge mode with index $n =77$ in the incomplete cell); (\textbf{b}) single-site excitation of the seventh waveguide from left to right at the top edge at $\gamma=1.0$; (\textbf{c}) single-site excitation of the ninth waveguide from top to bottom at the left edge at $\gamma=1.6$ ( overlapping with the edge mode with index $n =73$ in the complete cell).
}\label{figSI5}
\end{figure*}

\textbf{Supplementary Note 5. Single-site excitations at different array edges.}
Due to the $\mathcal{C}_3$ discrete rotational symmetry of the proposed HOTI structure, the resulting output patterns for excitations at the top edge of the structure are expected to match the output patterns obtained when the waveguides on the left side of the structure are excited. For $\gamma = 0.6$, the excitation of the waveguide in the incomplete unit cell at the top edge results in the beating of light predominantly scattered between three waveguides, as shown in Supplementary Figure \ref{figSI5}\textbf{a}, while it remains confined at the edge of the structure. As the energy of the input pulse increases, the tendency to concentrate the light in the excited waveguide becomes remarkable. The output patterns, in this case, show strong similarities with the output patterns obtained when the seventh waveguide was excited within the incomplete cell at the left edge of the structure (see Fig.~5\textbf{b} in the main text). The comparison of the output distributions for the excitation of the same waveguide at $\gamma=1.0$ in Supplementary Figure \ref{figSI5}\textbf{b} with the results from Fig.~4\textbf{b}, which show the excitation of the seventh waveguide in the same incomplete cell but at the left edge of the triangle, again illustrates the similarity of the diffraction patterns. Some differences in the local details of these patterns are due to the ellipticity of the waveguides and their orientation with respect to the upper and left edges of the triangle. These differences are usually only noticeable at $\gamma = 1.0$ when the diffraction into the bulk is significant. Finally, the results of the excitation of the ninth waveguide within the complete unit cell at the left edge of the array (see Supplementary Figure \ref{figSI5}\textbf{c}) show almost identical behavior to the excitation of a similar waveguide at the top edge of the array shown in Fig.~6\textbf{c} in the main text. These results, shown in Supplementary Figure \ref{figSI5}, confirm the high accuracy of the array fabrication, which leads to similar behavior of the excitations at different edges and in different corners, i.e., to an excellent reproducibility of the results.

\textbf{Supplementary Note 6. Two-site excitations for corner modes.}
As one can see from the structure of linear localized modes described above, some of the corner states are characterized by the presence of two dominant spots (in complete unit cells) that can be in-phase or out-of-phase. To efficiently excite such states in both experiment and theory, we have used two-site excitations. In the structure with trivial phase ($\gamma=1.0$), the output patterns for excitation with two out-of-phase beams, shown in Supplementary Figure \ref{figSI6}\textbf{a} and \ref{figSI6}\textbf{e}, exhibit strong diffraction into the bulk in the linear limit, which is gradually replaced by localization, but only when the input power exceeds a certain considerable threshold. In the array with $\gamma=1.6$, the same type of excitation leads to a highly efficient excitation of a corner mode with an index $n=74$ already in the linear regime (see Supplementary Figure \ref{figSI6}\textbf{b} and \ref{figSI6}\textbf{f}. In this case, increasing the input pulse energy leads to radiation into the bulk of the array (see columns two and three of Supplementary Figure \ref{figSI6}\textbf{b} and ~\ref{figSI6}\textbf{f}), and as soon as the power exceeds a certain threshold, the nonlinear output state is completely trapped in the input waveguides.

As described in the main text, at $\gamma=1.6$ there are two modes with indices $n=22$ and $121$ that represent bound states in the continuum (BICs), i.e., they are localized states with eigenvalues that belong to the bulk bands. These modes have in-phase or out-of-phase bright spots in the two outermost corner waveguides of the structure. In contrast, the corresponding modes at $\gamma=0.6$ and $\gamma=1.0$ are not bound (see lower panels of Fig.~2\textbf{d} in the main text). The excitation of mode $n=121$ with out-of-phase spots using two out-of-phase beams focused in two outermost corner waveguides (belonging to complete and incomplete unit cells) is shown in Figs.~4\textbf{d}, 5\textbf{d}, and 6\textbf{d} in the main text. Here we present the results for the excitation of mode $n=22$ with in-phase spots in the two outermost corner waveguides. They are shown in Supplementary Figure \ref{figSI6}\textbf{c} and \ref{figSI6}\textbf{d} for the displacement parameters $\gamma=1.0$ and $\gamma=1.6$ respectively. While with $\gamma=1.0$ in the trivial phase, localization is only achieved above a certain threshold pulse energy, with $\gamma=1.6$, such BIC is already excited at low pulse energies.
\begin{figure*}[htbp]
	\centering
	\includegraphics[width=1\columnwidth]{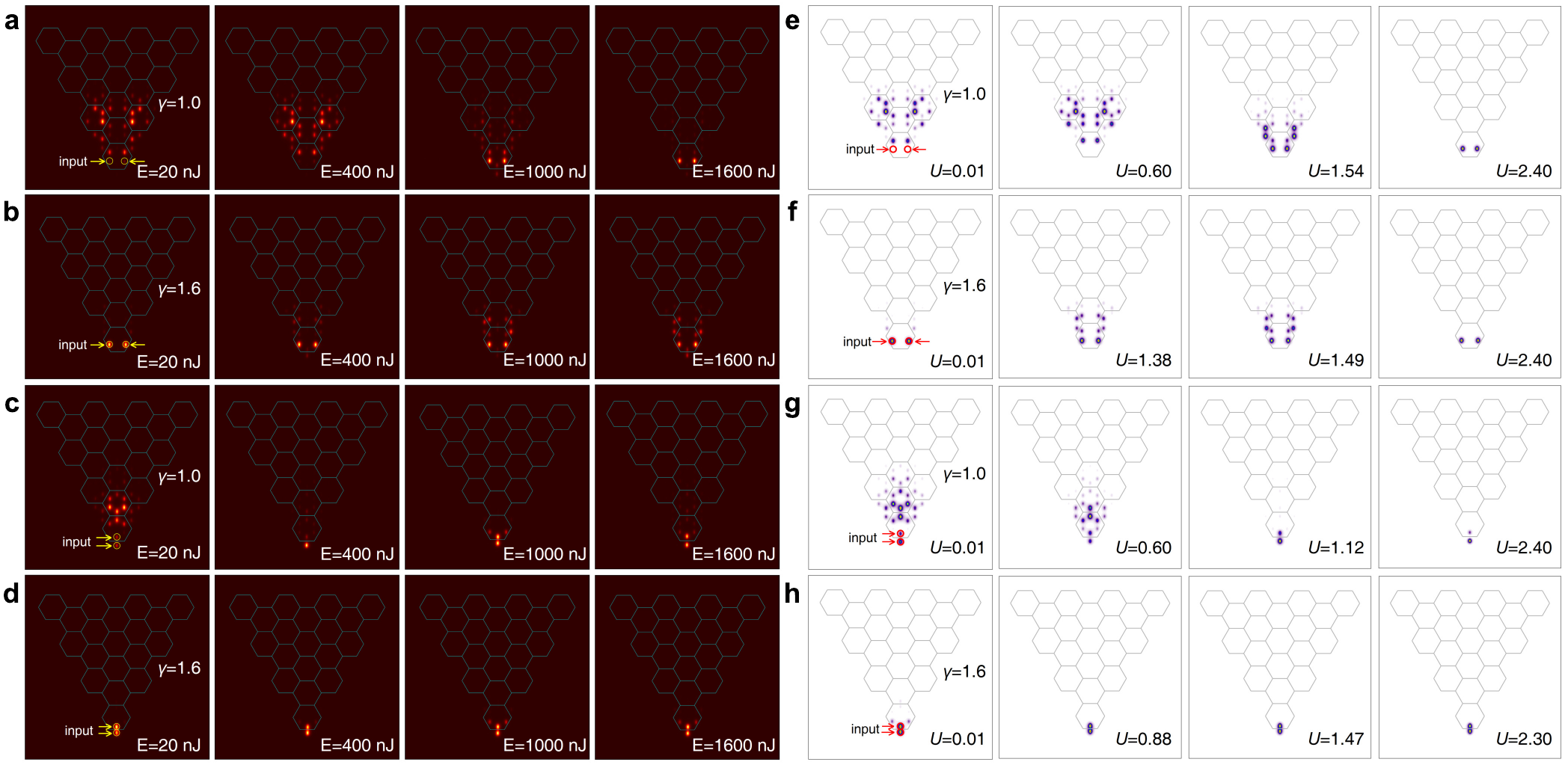}
	\caption{\textbf{Excitations of linear and nonlinear states using two-site inputs}. The arrangement and meaning of the panels are as in Supplementary Figure \ref{figSI5}. Out-of-phase excitation of two waveguides in a complete unit cell (overlap with corner mode with $n =74$) at $\gamma=1.0$ (\textbf{a}) and $\gamma=1.6$ (\textbf{b}); in-phase excitation of two bottom near-corner waveguides at $\gamma=1.0$ (\textbf{c}) and $\gamma=1.6$ (\textbf{d}), overlapping with the in-phase BIC mode with $n =22$.}
	\label{figSI6}
\end{figure*}